\documentclass[12pt,a4paper]{article}
\makeatletter
\newcommand{\vast}{\bBigg@{4}}
\newcommand{\Vast}{\bBigg@{5}}
\makeatother
\usepackage{amsmath}
\usepackage{amsfonts}
\usepackage{amssymb}
\usepackage{graphicx}
\usepackage{caption}
\usepackage{subcaption}
\usepackage{sidecap}
\usepackage{color}
\usepackage[colorlinks=true,allcolors=blue]{hyperref}
\usepackage{chngcntr}
\usepackage{cite}
\usepackage{authblk}
\author[1,2]{E D\'iaz-Bautista\footnote{ediazba@ipn.mx; ediaz@fis.cinvestav.mx}}
\author[3]{Y Betancur-Ocampo\footnote{ybetancur@icf.unam.mx}}
\affil[1]{\small Departamento de Formaci\'on B\'asica Disciplinaria, Unidad Profesional Interdisciplinaria de Ingenier\'ia Campus Hidalgo del Instituto Polit\'ecnico Nacional, Pachuca: Ciudad del Conocimiento y la Cultura, Carretera Pachuca-Actopan km 1+500, San Agust\'in Tlaxiaca, 42162 Hidalgo, M\'exico}
\affil[2]{\small Departamento de F\'isica, Cinvestav, P.O. Box. 14-740, 07000 Ciudad de M\'exico, M\'exico}
\affil[3]{\small Instituto de Ciencias F\'isicas, Universidad Nacional Aut\'onoma de M\'exico, 62210 Cuernavaca, M\'exico}
\title{Phase-space representation of Landau and electron coherent states for uniaxially strained graphene}
\date{ }

\begin{document}
	\maketitle
		\begin{abstract}
Recent experimental advances in the reconstruction of the Wigner function (WF) for electronic systems have led us to consider the possibility of employing this theoretical tool in the analysis of electron dynamics of uniaxially strained graphene. In this paper, we study the effect of strain on the WF of electrons in graphene under the interaction of a uniform magnetic field. This mechanical deformation modifies drastically the shape of the Wigner function of Landau and coherent states. The WF has a different behavior straining the material along the zigzag direction in comparison with the armchair one and favors the creation of electron coherent states. The time evolution of the WF for electron coherent states shows fluctuations between classical and quantum behavior around a closed path as time increases. The phase-space representation shows more clearly the effect of nonequidistant relative Landau levels in the time evolution of electron coherent states compared with other approaches. Our findings may be useful in establishing protocols for the realization of electron coherent states in graphene as well as a bridge between condensed matter and quantum optics.
	\end{abstract}

\section{Introduction}

Quantum systems are traditionally described by using a probabilistic interpretation of them. From this notion, the so-called wave function $\psi$ arises, with which we can calculate the probability of finding a particle in some space region by taking its square norm $\vert\psi\vert^2$ or the probability that a certain system is in one state or another. Applying the so-called evolution operator, we can know how the probabilities change over time \cite{Landau2}. However, this formulation about how to quantize a system is not unique mainly because many physicists try to explain the apparent controversy between the superposition principle of quantum mechanics and the probabilistic results of the measurement. As a consequence, there are a variety of formulations of quantum mechanics based on different interpretations of physical reality. Among such formulations, the phase-space representation places the position and momentum variables on equal footing \cite{w32,Cahill,berry77,bffls78,jsv81,igf82,barker83,hosw84,balazs84,takahashi86,royer91,lee95,Kenfack,h40,Case,Weinbub,Kryukov,Knight}. In this formulation of quantum mechanics, one leaves the wave-function idea to adopt the quasiprobability distribution and the operator action is replaced by a star product \cite{Kryukov}. Nowadays, the Wigner function (WF) constitutes one of the most important theoretical tools for describing quantum systems in the phase-space representation \cite{berry77,bffls78,jsv81,igf82,barker83,hosw84,balazs84,takahashi86,royer91,lee95,Weinbub}. This function widely appears in different physics and chemistry branches \cite{Weinbub}. Experimental reconstruction in quantum tomography experiments evidenced the significant role of the WF in quantum physics~\cite{Ding,Leiner,Gu,Knyazev,Ferraro,Marguerite}. In interference phenomena, the WF allows recognizing the pure states and their interacting parts, which is useful for studying quantum information processes~\cite{Birrittella,Rundle,Gonzalez}. Recently, there has been a significant increment in the employment of WF for electronic systems \cite{Jacoboni,Bauerle,Ferraro2,Morandi,Jullien,Mason,Iafrate,Ferry}, especially in the study of electron transport \cite{Morandi,Bauerle,Ferry}. These works exhibit that this quantum mechanics formulation connects different physics areas, treating indistinctly electrons and photons due to their wave nature. Likewise, interesting analogies between the effective Dirac-like approach depicting low-energy excitations in graphene and the Jaynes-Cummings model in quantum optics have been pointed out \cite{Mota,Ojeda,Jellal,Rusin,Dora,Goldman,Schliemann}.

On the other hand, the emergence of strain engineering has opened an important platform for exploiting the outstanding electronic, transport, and optical properties in graphene \cite{Pereira,Pereira2,Pereira3,Cocco,Ribeiro,Cadelano,Colombo,Castro,Stegmann,Stegmann2,Zhai,Peeters,Charlier,Settnes2,Betancur,Betancur2,Betancur3,Amorim,Naumis,Naumis2,Naumis3,Carrillo,Andrade,Barraza,Pellegrino,Vozmediano,Guinea,Arias,Sloan,Settnes,Gomes,Levy,Rechtsman,Ang,Choi,Naumov,Rostami,Deng,Shioya,Shioya2,Beams,Lee,Colangelo,Assili,Midtvedt,Haddad}. This topic started with the possibility of obtaining a gap opening \cite{Pereira,Cocco,Naumov}, but later multiple interesting effects were predicted such as the generation of pseudo magnetic fields \cite{Vozmediano,Guinea,Gomes} through the application of inhomogeneous strain \cite{Vozmediano,Guinea,Arias,Sloan,Settnes,Gomes,Levy,Rechtsman}, and the modulation of physical properties without detriment to the Dirac cones in a wide deformation range, giving rise, for instance, to the contraction of Landau levels (LLs) \cite{Betancur}, optical absorption \cite{Pereira3,Pellegrino,Naumis3}, and magnetostrain-driven quantum heat engines \cite{Munoz}. In this context, the realization of exotic ballistic transport phenomena was also predicted as the collimation effect of electrons \cite{Pereira2}, valley beam splitters \cite{Stegmann2,Zhai,Peeters,Charlier,Settnes2}, Klein tunneling deviation, and asymmetric Veselago lenses \cite{Betancur2}. These inspirational works motivate us to introduce the Wigner matrix (WM) formalism for uniaxially strained graphene under the presence of a uniform magnetic field, to describe the electron dynamics in the phase-space representation. The recent advances about the quantum tomography of an electron \cite{Jullien} might stimulate the experimental reconstruction of the WF in two-dimensional materials, particularly for uniaxially strained graphene due to the existing important experimental contributions so far \cite{Shioya,Shioya2,Beams,Lee,Colangelo}. Moreover, the resemblance of electron and light quantum optics is more evidenced by the WF analysis.

For these motivations, we propose the study of quasi-probability distributions for describing the time evolution of electron dynamics in uniaxially strained graphene under a uniform and perpendicular magnetic field. We start by developing an effective Dirac-Weyl (DW) model based on the tight-binding (TB) approach to nearest neighbors around a Dirac point. We relate the geometrical parameters of the elliptical Dirac cones with the components of the strain tensor, showing a good agreement with recent density functional theory (DFT) calculations \cite{Betancur}. Such parameters allow us to modulate successfully the shape of the WF through the tensile strain. We calculate the Landau states from our effective model. Using the integral representation of the WF, we determine the components of the WM as a function of the tensile strain and the magnitude of the magnetic field. We also build electron coherent states from the Landau quantization through the eigenstates of annihilation operators of pseudo-spin-1/2 systems and obtain the corresponding WM. We find that stretching along the zigzag ($\mathcal{Z}$) direction favors the observation of electron coherent states. From the time evolution of the WM of electron coherent states, we identify aspects of electron classical motion in a uniform magnetic field as well as evidence of fluctuations in the quasi-probability distribution, where negative values indicate a nonclassical behavior of these states. At the time this paper has been developed, the analysis of time evolution of electron dynamics in phase space for strained two-dimensional materials has not yet been performed.

\section{Effective tight-binding Hamiltonian in uniaxially \\ strained graphene}

\begin{figure}
	\centering
\begin{tabular}{ccc}
(a) \qquad \qquad \qquad \qquad \qquad \qquad & (b) \qquad \qquad \qquad \qquad \qquad \qquad & (c) \qquad \qquad \qquad \qquad \qquad \qquad \\
	\includegraphics[trim = 0mm 0mm 0mm 0mm, scale= 0.12, clip]{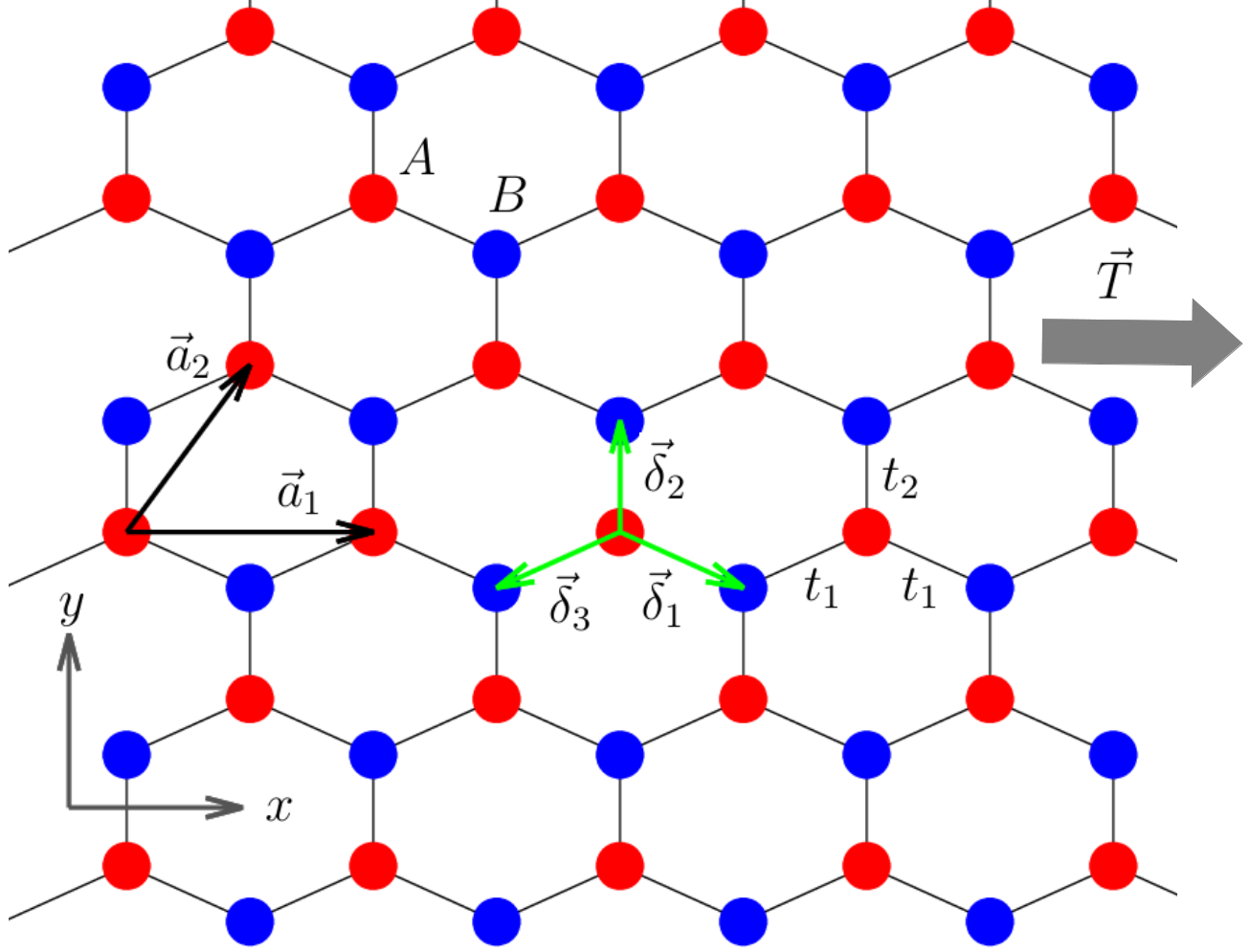} &
\includegraphics[trim = 0mm 0mm 0mm 0mm, scale= 0.17, clip]{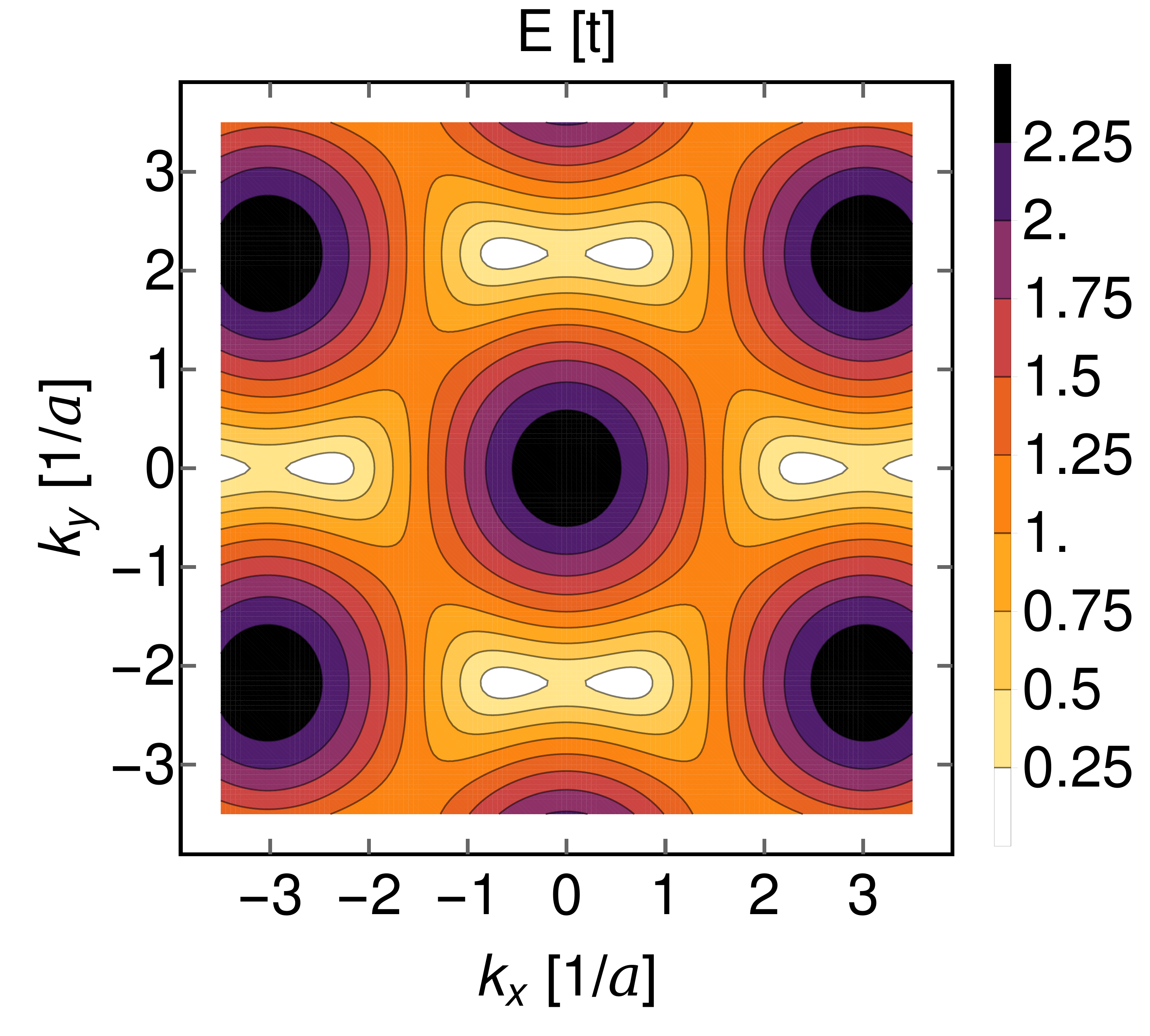} &
\includegraphics[trim = 0mm 0mm 0mm 0mm, scale= 0.17, clip]{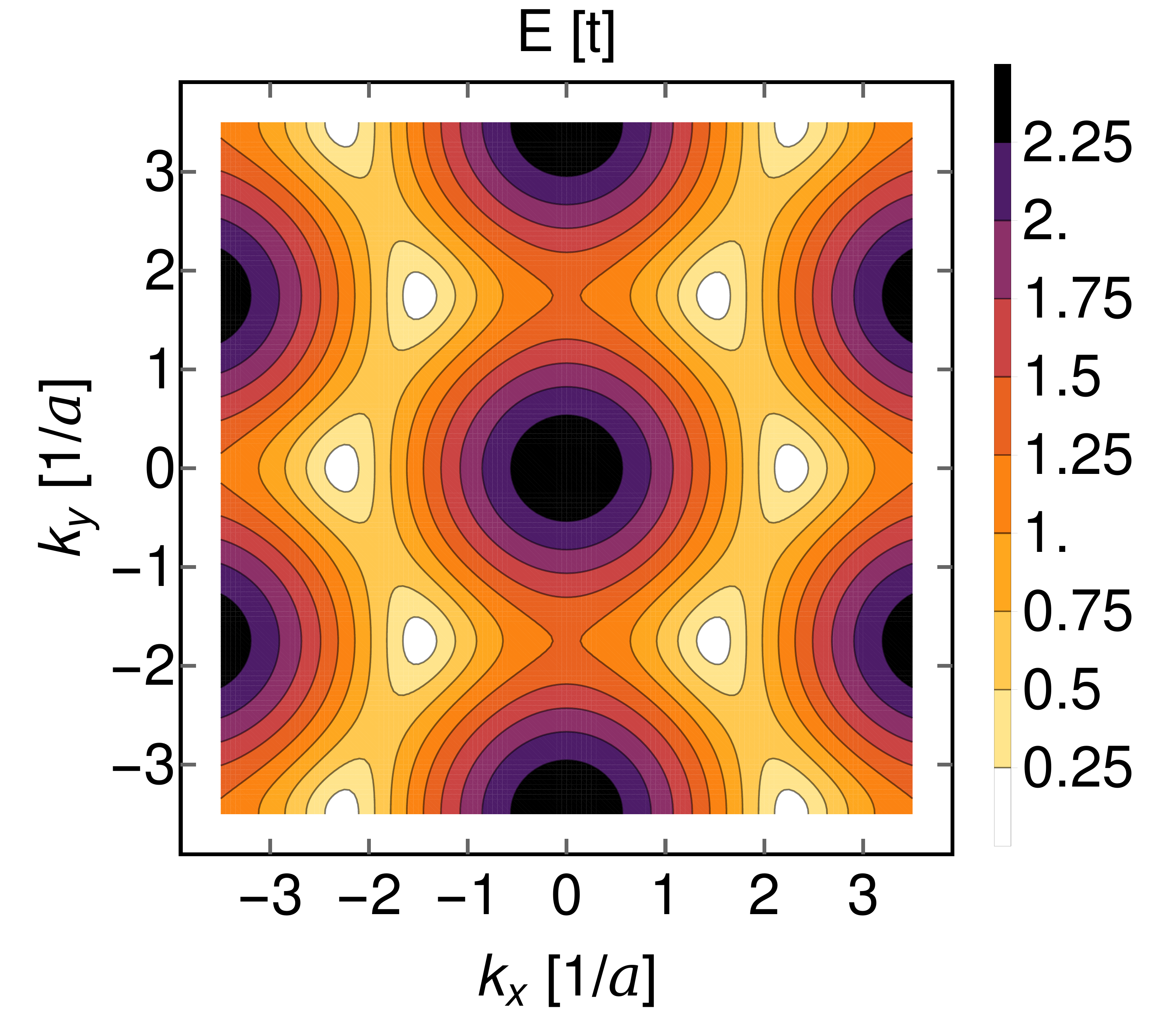}
\end{tabular}
	\caption{(a) Schematic representation of uniaxially strained graphene where red and blue circles indicate the sites of triangular sublattices A and  B, respectively. Uniaxial strain along the zigzag or armchair direction has two different nearest-neighbor hopping parameters $t_1$ and $t_2$. The lattice vectors are denoted by $\vec{a}_1$ and $\vec{a}_2$. (b) and (c) Energy contour of the conduction band near the first Brillouin zone for deformation of $20 \%$ along the zigzag and armchair direction, respectively.}
	\label{system}
\end{figure}

\subsection{Tight-binding approach to nearest neighbors}

For describing the WM in strained graphene, we start using the TB approach to an anisotropic hexagonal lattice \cite{Pereira,Naumis,Midtvedt,Betancur2,wallace47}. In this approach, the $p_z$ orbital of the carbon atom is decoupled with the $\sigma$ ones. Other approximations consist of neglecting the overlap between $p_z$ orbitals and second nearest-neighbor interactions. Since there are two atoms per unit cell and we assume an infinity extended layer, the TB is simplified considerably and reduced to a model of two energy bands \cite{Goerbig,Goerbig2,Betancur2}. This effective model only depends on two hopping parameters $t_1$ and $t_2$ for quinoid type deformations, which quantify the probability amplitude that an electron hops to the nearest atom. Most of the properties of strained graphene depend strongly on the value of these parameters. For instance, an excessive value of one of them can cause a gap opening \cite{Pereira,Cocco}. These parameters are related directly to the bond lengths through an exponential decay rule. With the application of a uniaxial tension $T$ to graphene, the atomic sites are displaced and modify the hopping parameter values, as shown in Fig.~\ref{system}. The positions of nearest neighbors are denoted by $\vec{\delta_1}$, $\vec{\delta_2}$, and $\vec{\delta_3}$ on the underlying sublattice A, where the Cartesian system is set with the $x$ axis along with the $\mathcal{Z}$ bond. The lattice vectors $\vec{a}_1$ and $\vec{a}_2$ allow connecting the whole positions in the deformed hexagonal lattice. From elasticity theory \cite{Pereira,Colombo,Cadelano,Landau}, the uniaxial strain tensors along $\mathcal{Z}$ and armchair ($\mathcal{A}$) directions are, respectively,
\begin{equation}\label{str}
{\bf \epsilon}_{\mathcal{Z}} = \left(\begin{array}{cc} 
1 & 0\\
0 & -\nu
\end{array}\right)\epsilon, \qquad {\bf \epsilon}_{\mathcal{A}} = \left(\begin{array}{cc} 
-\nu & 0\\
0 & 1
\end{array}\right)\epsilon,
\end{equation}
where $\nu$ is the Poisson ratio of graphene \cite{Pereira2,Cadelano} and the tensile strain $\epsilon$ is proportional to the magnitude of tension $T$. This quantifies the percentage of deformation \cite{Pereira}. According to Eq.~(\ref{str}) and applying positive (negative) deformation $\epsilon$ along a particular axis, the perpendicular direction contracts (expands) by $-\nu\epsilon$. The atomic positions in uniaxially strained graphene are $\vec{r} = (\mathbb{I} + \bar{\epsilon})\vec{r}_0$, where $\vec{r}_0$ indicates the sites on the pristine graphene, $\mathbb{I}$ denotes the $2\times2$ unity matrix, and $\bar{\epsilon}$ is the deformation tensor. Thus, the deformed lattice vectors for uniaxial strain in the $\mathcal{Z}$ direction are
\begin{subequations}\label{lattvszz}
	\begin{eqnarray}
	\vec{a}^{\mathcal{Z}}_1 & = & \sqrt{3}a_0\hat{x}(1 + \epsilon), \\
	\vec{a}^{\mathcal{Z}}_2 & = & \frac{\sqrt{3}}{2}a_0[\hat{x}(1 + \epsilon) + \sqrt{3}\hat{y}(1 -\nu\epsilon)],
	\end{eqnarray}
\end{subequations}
while for the $\mathcal{A}$ direction they are
\begin{subequations}\label{lattvsac}
	\begin{eqnarray}
	\vec{a}^{\mathcal{A}}_1 & = & \sqrt{3}a_0\hat{x}(1 - \nu\epsilon), \\
	\vec{a}^{\mathcal{A}}_2 & = & \frac{\sqrt{3}}{2}a_0[\hat{x}(1 - \nu\epsilon) + \sqrt{3}\hat{y}(1 +\epsilon)],
	\end{eqnarray}
\end{subequations}
where $a_0$ is the bond length in pristine graphene \cite{Castro}. Since the nearest-neighbor sites are given by $\vec{\delta}_1 = 2\vec{a}_1/3 - \vec{a}_2/3$, $\vec{\delta}_2 = 2\vec{a}_2/3 - \vec{a}_1/3$, and $\vec{\delta}_3 = -\vec{\delta}_1 -\vec{\delta}_2$, they are also related with the tensile strain.

To obtain an effective model based on the TB approach, we develop the following TB Hamiltonian in the Fourier basis from a plane-wave ansatz \cite{Betancur2,wallace47,Goerbig}
\begin{equation}
H_{\text{TB}}^{K} = \sum^3_{j = 1}\left[\begin{array}{cc}
0 & t_j\textrm{e}^{i\vec{k}\cdot\vec{\delta}_j}\\
t_j\textrm{e}^{-i\vec{k}\cdot\vec{\delta}_j} & 0
\end{array}\right].
\label{H}
\end{equation} 
The hopping parameters $t_j$ can be modeled using an exponential decay rule $t_j = t\textrm{exp}[-\beta(\delta_j/a - 1)]$, where $\beta$ is the Gr\"uneisen constant, $t$ is the hopping in pristine graphene, and $\delta_j$ are the deformed bond lengths \cite{Pereira,Castro,Betancur2,Papas}. The relation of $t_j$ as a function of strain parameters is completed when we express the deformed lengths in terms of tensile strain, which for $\mathcal{Z}$ deformations are given by
\begin{subequations}\label{dzz}
	\begin{align}
	\delta^{\mathcal{Z}}_1 & =  \delta^{\mathcal{Z}}_3= a_0\sqrt{\left[1 + \frac{1}{4}(3 - \nu)\epsilon\right]^2 + \frac{3}{16}(1+ \nu)^2\epsilon^2}, \\
	\delta^{\mathcal{Z}}_2 & =  a_0(1 -\nu\epsilon),
	\end{align}
\end{subequations}
while for uniaxial strain along the $\mathcal{A}$ direction
\begin{subequations}\label{dac}
	\begin{align}
	\delta^{\mathcal{A}}_1 & =  \delta^{\mathcal{A}}_3= a_0\sqrt{\left[1 + \frac{1}{4}(1 - 3\nu)\epsilon\right]^2 + \frac{3}{16}(1+ \nu)^2\epsilon^2}, \\
	\delta^{\mathcal{A}}_2 & =  a_0(1 + \epsilon).
	\end{align}
\end{subequations}
The electronic band structure of uniaxially strained graphene is obtained from the eigenenergies of the Hamiltonian \eqref{H}
\begin{equation}
E_s(\vec{k}) = s\left|\sum^3_{j = 1}t_j\textrm{e}^{-i\vec{k}\cdot\vec{\delta}_j}\right|,
\end{equation}
where the band index $s = 1\,(-1)$ corresponds to the conduction (valence) band. The energy bands can be modified applying uniaxial strain, as shown in Fig. \ref{system}(b) and (c). DFT and TB calculations have an excellent agreement within the low-energy regime \cite{Ribeiro}. For that reason, the TB Hamiltonian \eqref{H} is expanded around the Dirac point performing $\vec{k} = \vec{q} + \vec{K}_D$ and the Dirac point position $\vec{K}_D$ satisfies $\sum_jt_j\textrm{exp}(-i\vec{K}_D\cdot\vec{\delta}_j) = 0$. Thus, the effective Dirac-Weyl-like Hamiltonian in the continuum approximation is
\begin{equation}\label{HD}
H_{\text{D}}=v_{\rm F}\left[\begin{array}{cc}
0 & ap_x -ibp_y\\
ap_x + ibp_y & 0
\end{array}\right],
\end{equation}
where the quantities  
\begin{subequations}
	\begin{align}
a & = \frac{2}{3}\sum^3_{j = 1}\frac{\delta_{jx}}{a_0}\frac{t_j}{t}\sin(\vec{K}_D\cdot\vec{\delta}_j),\\
b & = \frac{2}{3}\sum^3_{j = 1}\frac{\delta_{jy}}{a_0}\frac{t_j}{t}\cos(\vec{K}_D\cdot\vec{\delta}_j).
\label{w}
\end{align}
\end{subequations} 
are expressed as functions of the lattice vectors and hopping parameters. Taking into account the relation 
\begin{equation}
\cos[\vec{K}_D\cdot(\vec{\delta}_1 - \vec{\delta}_2)] = -\frac{t_2}{2t_1},
\end{equation}
which is obtained from the Dirac points equation $\sum^3_jt_j\textrm{e}^{-i\vec{K}_D\cdot\vec{\delta}_j} = 0$, it is possible to show that
\begin{eqnarray}
a & = & \frac{2}{3a_0t}\sqrt{a^2_{1x}t^2_1 + (a_{2x} - a_{1x})a_{2x}t^2_2}, \nonumber\\
b & = & \frac{2}{3a_0t}\sqrt{a^2_{1y}t^2_1 + (a_{2y} - a_{1y})a_{2y}t^2_2}.
\label{ayb}
\end{eqnarray}

\begin{figure}
	\centering
\begin{tabular}{ccc}
(a) \qquad \qquad \qquad \qquad \qquad \qquad & (b) \qquad \qquad \qquad \qquad \qquad \qquad & (c) \qquad \qquad \qquad \qquad \qquad \qquad \\
	\includegraphics[trim = 0mm 0mm 0mm 0mm, scale= 0.17, clip]{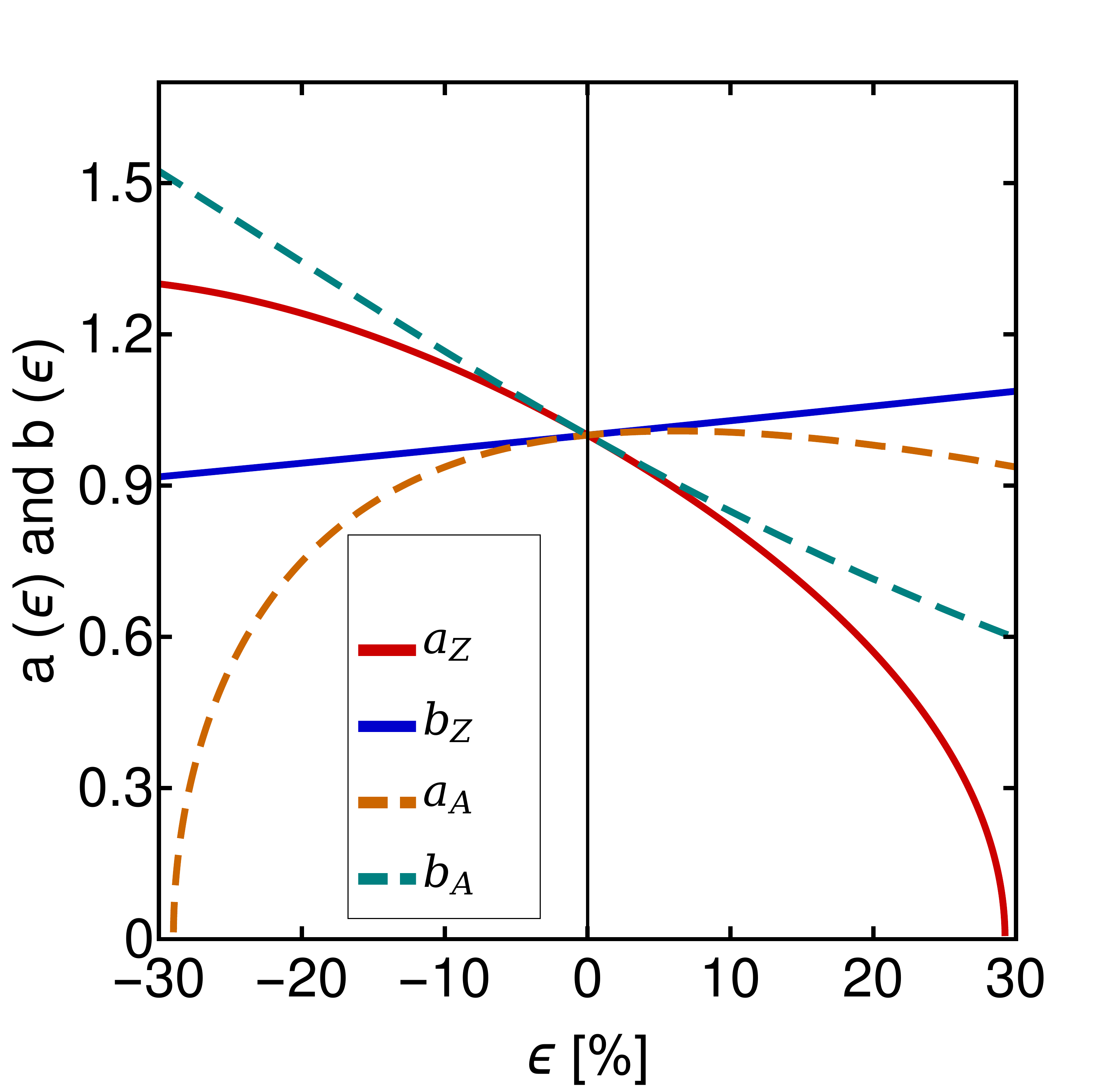} &
\includegraphics[trim = 0mm 0mm 0mm 0mm, scale= 0.17, clip]{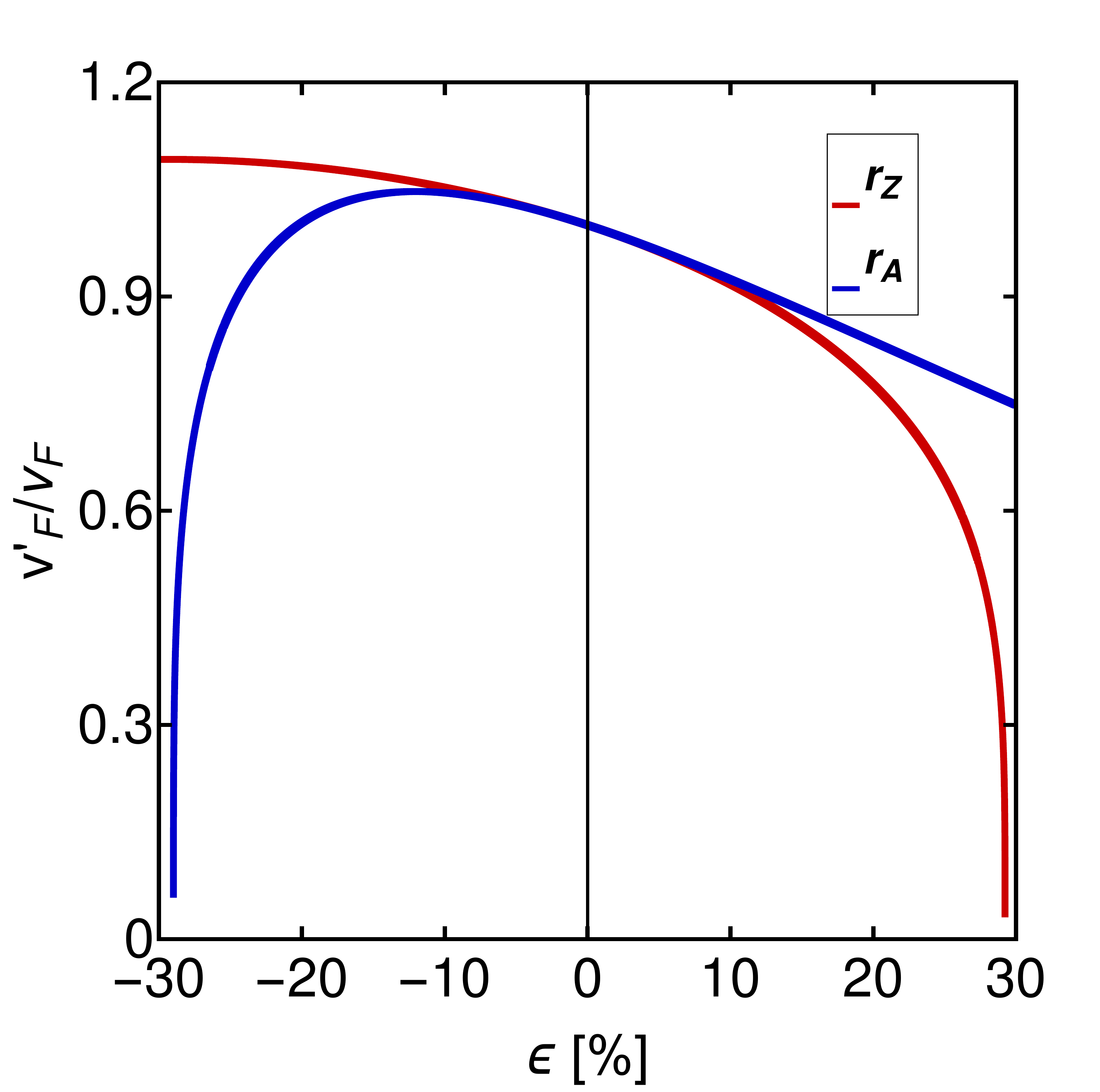}&
\includegraphics[trim = 0mm 0mm 0mm 0mm, scale= 0.17, clip]{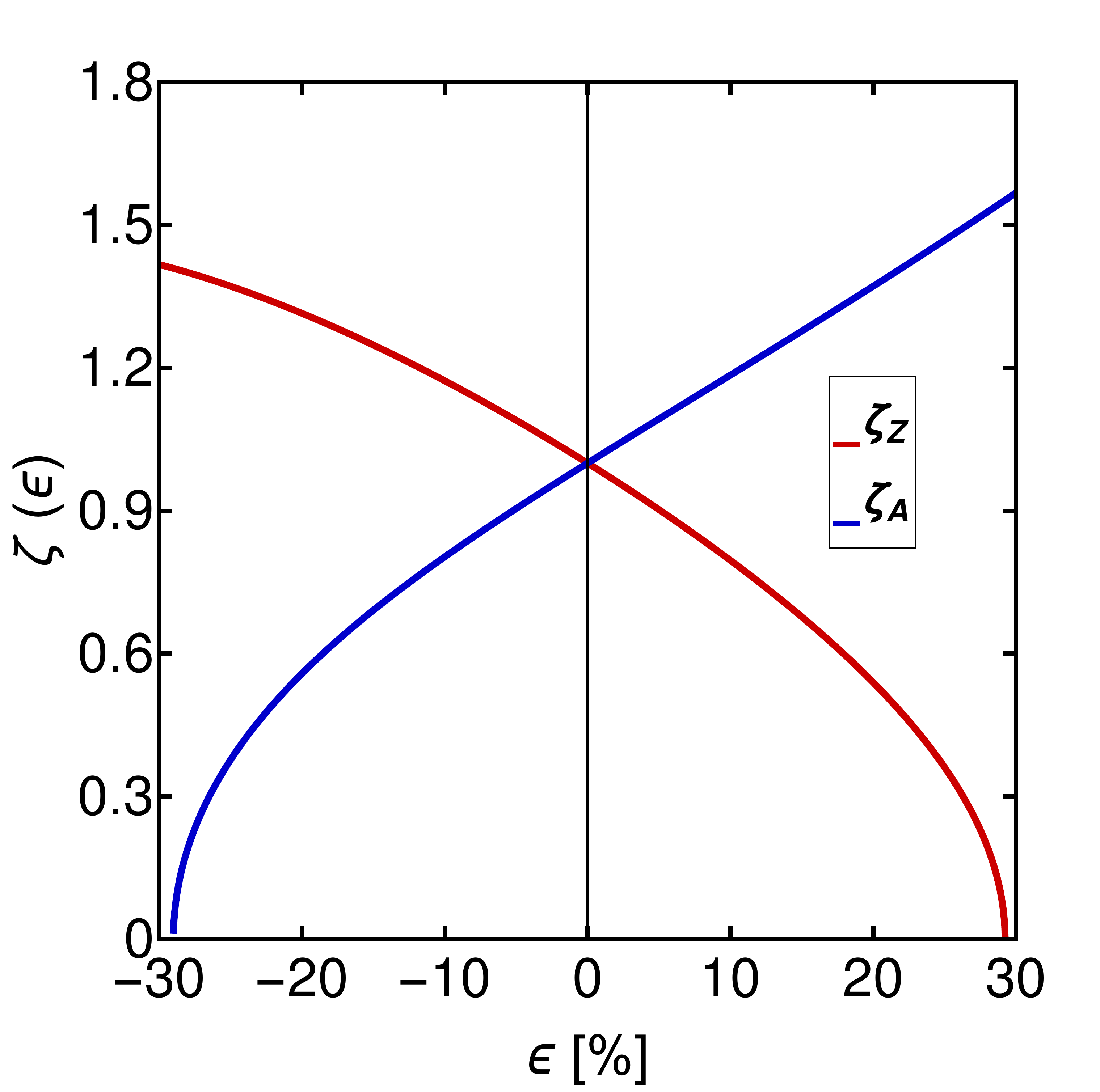}
\end{tabular}
	\caption{Geometrical parameters $a$ and $b$ (a), effective Fermi velocity $v'_{\rm F} = v_{\rm F}\sqrt{ab}$ (b), and supersymmetric potential parameter $\zeta = a/b$ (c) as a function of the tensile strain parameter $\epsilon$ along the zigzag and armchair directions.}
	\label{GA}
\end{figure}

It is important to mention that the parameters $a$ and $b$ can be related to the extremal angle of the elliptical Dirac cones \cite{Betancur}. This shows clearly that the physical properties of strained graphene can be modulated by the geometrical parameters of the Dirac cone through the strain using the relation \eqref{ayb}. These quantities depend on the strain direction by evaluating the corresponding expressions of lattice vectors \eqref{lattvszz} and \eqref{lattvsac} and bond lengths \eqref{dzz} and \eqref{dac} for $\mathcal{Z}$ and $\mathcal{A}$ directions, respectively. In the geometrical approach, the parameters $a$ and $b$ are fitted to the energy bands obtained from DFT calculations \cite{Betancur}. The behavior of $a$ and $b$ is similar to the elastic strain observed from the geometrical approach \cite{Betancur} (see Fig. \ref{GA}(a)). We can observe that for strain values up to 10\% the geometrical parameters satisfy $a_{\mathcal{Z}} \approx b_{\mathcal{A}}$ and $b_{\mathcal{Z}} \approx a_{\mathcal{A}}$. Beyond this range, these geometrical parameters are different. This behavior is following the stress-strain relationship of graphene \cite{Cadelano}; the elastic response is isotropic (anisotropic) and linear (nonlinear) for $\epsilon < 10 \%$ ($\epsilon > 10 \%$). Therefore, the analytical formulas \eqref{ayb} indicate how the strain affects the geometrical parameters $a$ and $b$ as well as other electronic properties, such as effective Fermi velocity $v'_{\rm F} = v_{\rm F}\sqrt{ab}$ (see Fig. \ref{GA}(b)) and the supersymmetric potential parameter $\zeta = a/b$ (see Fig. \ref{GA}(c)). Likewise, we will show the effect of strain on the WF of Landau and coherent states in forthcoming sections.

\subsection{Dirac-Weyl equation under uniaxial strain}\label{dirac}
The effective Dirac-Weyl Hamiltonian \eqref{HD} has validity in the energy range of 0-0.3 eV, where DFT and TB calculations agree very well \cite{Ribeiro}. It is important to mention that there exist other models for depicting the electronic properties of strained graphene using a pseudo-vector potential. Such theoretical works have been useful for studying graphene under inhomogeneous deformations \cite{Amorim}. However, in the uniform strain case it has been shown that those models do not describe correctly the electron dynamics because the high-symmetry point $K$ in the first Brillouin zone is far away from the Dirac point, where the dispersion relation can be nonlinear \cite{Naumis}. For that reason, an effective Dirac-Weyl model around the Dirac point is more adequate for studying the physical properties of uniaxially strained graphene instead of one in the high-symmetry points. Moreover, this model constitutes a simple way to promote the Fermi velocity to a tensor character preventing the generation of pseudomagnetic fields by the homogeneity of the deformation. An analogous procedure can be implemented at the $K'_D$ valley to obtain the corresponding effective Dirac-like Hamiltonian.

Now, let us consider the DW equation under uniform uniaxial strain
	\begin{equation}\label{WDE}
	H\Psi(x,y)=v_{\rm F}(a\sigma_{x}\pi_x+b\sigma_{y}\pi_y)\Psi(x,y)=E\Psi(x,y),
	\end{equation}
	where the geometrical parameters $a$ and $b$ are given by Eq.~(\ref{ayb}) and the linear momentum in the presence of a magnetic field and under the Peierls substitution is $\pi_{x,y}=p_{x,y}+eA_{x,y}$, with $\vec{p}$ being the kinetic momentum in free field and $\vec{A}$ the vector potential that generates a magnetic field aligned perpendicularly to the graphene sample. In a Landau-like gauge,
	\begin{equation}
	\vec{A}(x,y)=A_y(x)\hat{y}, \quad \vec{B}=\nabla\times\vec{A}=B(x)\hat{z},
	\end{equation}
the linear momentum $p_y = \hbar k$, with $k$ the wave vector in $y$, is conserved and therefore we can write the pseudo-spinor ansatz as
	\begin{equation}
	\Psi(x,y)=\exp(iky)\left(\begin{array}{c}
	\psi^+(x) \\
	\psi^-(x)
	\end{array}\right).\label{spinor}
	\end{equation}
	Substituting~(\ref{spinor}) into~(\ref{WDE}), two coupled equations arise, namely
		\begin{equation}
	\left[ap_x\pm ib\left(k\hbar+eA_y(x)\right)\right]\psi^\pm(x)=\frac{E}{v_{\rm F}}\psi^\mp(x).
	\end{equation}
	These equations are decoupled to obtain
		\begin{equation}
	\left[-\zeta\frac{d^2}{dx^2}+V^\pm_\zeta(x)\right]\psi^\pm(x)=\varepsilon^{\pm2}\psi^\pm(x),
	\label{DS}
	\end{equation}
	where $\varepsilon^\pm=E/(v_{\rm F}\hbar\sqrt{ab})$ and
	\begin{equation}
	V^\pm_\zeta(x)=\zeta^{-1}\left(k+\frac{eA_y(x)}{\hbar}\right)^2\pm\frac{e}{\hbar}\frac{dA_y(x)}{dx}.
	\end{equation}
This effective one-dimensional potential has been discussed in a supersymmetric point of view \cite{Concha} and we show how this potential through the parameter $\zeta$ is tuned with the tensile strain $\epsilon$ in Fig. \ref{GA}(c). In the case of an inhomogeneous magnetic field, there is a little number of vector potential profiles that allow obtaining exact and analytical solutions of energy-level spectra and wave functions of the electron \cite{Concha,Kuru}. In general, it is necessary to solve numerically the Sturm-Liouville problem given by the decoupled systems \eqref{DS}. For a uniform magnetic field $\vec{B}=B\hat{z}$ we can obtain an exact solution using the Landau gauge $\vec{A}=Bx\hat{y}$. In this way, the Hamiltonians and effective potential in the supersymmetric approach are, respectively,
	\begin{equation}\label{13}
	H^\pm_\zeta=-\zeta\frac{d^2}{dx^2}+V^\pm_\zeta(x), \quad V^\pm_\zeta(x)=\frac{\zeta\omega^2_\zeta}{4}\left(x+\frac{2k}{\zeta\omega_{\zeta}}\right)^2\pm\frac{\zeta\omega_\zeta}{2},
	\end{equation}
where $\omega_\zeta$ is a frequency defined by $\omega_\zeta=\frac{2eB}{\zeta\,\hbar}$. Since this problem is very similar to solve the quantum harmonic oscillator, the LL spectra are straightforwardly obtained \cite{Betancur}
	\begin{equation}\label{LLs}
         E_{n} = s v'_{\rm F} \sqrt{2ne\hbar B},
        \end{equation}
where $v'_{\rm F}=v_{\rm F}\sqrt{ab}$ is the effective Fermi velocity. The Landau-level index $n$ runs over $0, 1, 2, ...$ and the positive (negative) energy corresponds to electrons in the conduction (valence) band, as given by the band index $s$. When one performs the same analysis in the other elliptical Dirac cone $K'_D$, the LL spectra are very similar to those shown in \eqref{LLs}. However, each valley contributes differently for the Landau state with $n = 0$ \cite{Goerbig}, namely, the zeroth Landau state receives a contribution of the sublattice A (B) in the valley $K_D$ ($K'_D$). This important fact gives rise to the anomalous quantum Hall effect \cite{Novoselov,Zhang,Andrei}. Furthermore, the quantity $\sqrt{ab}$ is always less than 1 for positive deformations (see Fig. \ref{GA}(b)), causing the contraction of LL spectra \cite{Betancur,Sahalianov}. In contrast, for compressing a graphene lattice the quantity $\sqrt{ab}$ increases and we can expand the LL spectra. It is important to mention that the LLs in Eq. \eqref{LLs} are only valid for low magnetic fields and energies \cite{Goerbig,Goerbig2}. Such an observation is useful if one considers building electron coherent states in uniaxially strained graphene since the largest possible number of LLs is needed within the energy range, where the dispersion relation is linear. In this way, the most convenient route for a tentative developing of electron coherent states is stretching graphene along the $\mathcal{Z}$ direction and using low magnetic fields. As seen in Fig. \ref{GA}(b), the Fermi velocity is lower in the $\mathcal{Z}$ direction than in the $\mathcal{A}$ direction for positive deformations and therefore the LL spectra are compressed more efficiently.
	
	The exact solution for the wave function of electrons in strained graphene with a uniform magnetic field turns out to be
\begin{equation}\label{17}
	\Psi_{n}(x,y)=\frac{\exp\left(iky\right)}{\sqrt{2^{(1-\delta_{0n})}}}\left(\begin{array}{c}
	(1-\delta_{0n})\psi_{n-1}(x) \\
	i\,\lambda s\,\psi_{n}(x)
	\end{array}\right),
	\end{equation}
and the components of the pseudo-spinor are given by
	\begin{equation}\label{5}
	\psi_n(x)=\sqrt{\frac{1}{2^nn!}\left(\frac{\omega_\zeta}{2\pi}\right)^{1/2}}\exp\left[-\frac{\omega_\zeta}{4}\left(x+\frac{2k}{\zeta\omega_\zeta}\right)^2\right]H_n\left[\sqrt{\frac{\omega_\zeta}{2}}\left(x+\frac{2k}{\zeta\omega_\zeta}\right)\right].
	\end{equation}
The quantity $\delta_{mn}$ denotes the Kronecker delta, $\psi^-_n\equiv\psi_{n}$, $\psi^+_n\equiv\psi_{n-1}$ and $\lambda=+/-$ describes Dirac fermions at the $K_D/K'_D$ valley. It is worth remarking that the wave function in the upper (lower) component of the pseudo-spinor (\ref{17}) corresponds to an electron at the $K_D$ valley in the sublattice A (B), while at the $K'_D$ valley the roles of upper and lower components are interchanged. These wave functions will be used for determining the coherent states and WM from an integral representation. In all our calculations, we set the following physical constants $\hbar = e = 1$. Also, without loss of generality in the next sections, we will focus on the conduction band ($s=1$).

\section{Electron coherent states of strained graphene}

\subsection{Annihilation operator}\label{annihilation}
First, for obtaining coherent states in electronic systems, we introduce the definition of creation and annihilation operators for electrons of strained graphene in the presence of a uniform magnetic field. Thus, let us define the following dimensionless differential operators
\begin{equation}
\theta^\pm=\frac{1}{\sqrt{2}}\left(\mp\frac{d}{d\xi}+\xi\right),\quad \theta^+=\left(\theta^-\right)^\dagger, \quad \xi=\sqrt{\frac{\omega_\zeta}{2}}\left(x+\frac{2k}{\zeta\omega_\zeta}\right),
\end{equation}
that satisfy the commutation relation
\begin{equation}
[\theta^-,\theta^+]=1.
\end{equation}
This relation implies that the set of operators $\{\theta^+,\theta^-,1\}$ generates a Heisenberg-Weyl (HW) algebra \cite{Diaz,Diaz2,Diaz3}.

Now, the action of the operators $\theta^\pm$ on the eigenfunctions $\psi_n$ in \eqref{5} is
\begin{equation}
\theta^-\psi_n=\sqrt{n}\psi_{n-1}, \quad \theta^+\psi_n=\sqrt{n+1}\psi_{n+1},
\end{equation}
so that $\theta^-$ ($\theta^+$) is the annihilation (creation) operator. In terms of these ladder operators, we can define the following dimensionless $2\times2$ Hamiltonian $\mathcal{H}$ that acts on the $x$-dependent pseudo-spinors in Eq. \eqref{17}
\begin{equation}
\mathcal{H}=\left[\begin{array}{cc}
0 & -i\theta^- \\
i\theta^+ & 0
\end{array}\right].
\end{equation}

To build coherent states in graphene, one can define a generalized annihilation operator $\Theta^-$ as
\begin{equation}\label{opanh}
\Theta^{-}=\left[\begin{array}{c c}
\cos(\delta)\frac{\sqrt{N+2}}{\sqrt{N+1}}\theta^- & \lambda\sin(\delta)\frac{1}{\sqrt{N+1}}(\theta^-)^2 \\
-\lambda\sin(\delta)\sqrt{N+1} & \cos(\delta)\theta^-
\end{array}\right], \quad \Theta^+=(\Theta^-)^\dagger,
\end{equation}
such that
\begin{equation}
\Theta^-\Psi_{n}(x,y)=\frac{\exp(i\,\delta)}{\sqrt{2^{\delta_{1n}}}}\sqrt{n}\Psi_{n-1}(x,y), \quad n=0,1,2,\dots,
\end{equation}
where $N=\theta^+\theta^-$ is the number operator. The parameter $\delta\in[0,2\pi]$ in Eq.~(\ref{opanh}) allows us to consider either a diagonal or nondiagonal form for such an operator in order to mix, exchange, or not mix or exchange the components that belong to different sublattices (A or B) in the pseudo-spinor.

On the other hand, the operators $\Theta^\pm$ satisfy the following commutation relation
\begin{equation}
[\Theta^-,\Theta^+]=\mathbb{I},
\end{equation}
that also generates the HW algebra.

\subsection{Coherent states as eigenstates of $\Theta^-$}
We obtain the coherent states $\Psi_{\alpha}(x,y)$ as eigenstates of the operator $\Theta^-$:
\begin{equation}
\Theta^-\Psi_{\alpha}(x,y)=\alpha\Psi_{\alpha}(x,y), \quad \alpha\in\mathbb{C},
\end{equation}
where $\alpha$ is the eigenvalue of $\Theta^-$. The $|\alpha|^2$ indicates the average electron number, while the $\alpha$ phase is the polar angle in phase space. The electron coherent states are expanded in the Landau states basis \eqref{5} as
\begin{equation}
\Psi_{\alpha}(x,y)=a_0\Psi_{0}(x,y)+\sum_{n=1}^{\infty}a_n\Psi_{n}(x,y).
\end{equation}

Upon inserting these states into the corresponding eigenvalue equation, we get the following relations
\begin{equation}
a_1=\sqrt{2}\tilde{\alpha}a_0, \quad a_{n+1}\sqrt{n+1}=\tilde{\alpha}a_n,
\end{equation}
where $\tilde{\alpha}=\alpha\exp\left(-i\,\delta\right)$. This means that the $\delta$ parameter translates as a phase factor, that will affect the eigenvalue $\alpha$, and $a_0$ is a free parameter that can be determined by the normalization process.

After straightforward calculations, the coherent states turn out to be \cite{Diaz2}:
\begin{eqnarray}
\Psi_{\alpha}(x,y)& = & \frac{1}{\sqrt{2\exp(|\tilde{\alpha}|^2) -1}}\left[\Psi_0(x,y)+\sum^{\infty}_{n=1}\frac{\sqrt{2}\tilde{\alpha}^n}{\sqrt{n!}}\Psi_n(x,y)\right]\nonumber\\
 & = & \frac{e^{iky}}{\sqrt{2\exp(|\tilde{\alpha}|^2) -1}}\left(\begin{array}{c}
\psi'_{\alpha}(x)\\
i\psi_{\alpha}(x)
\end{array}\right),
\label{CE}
\end{eqnarray}
where
\begin{subequations}
	\begin{align}
	\psi'_{\alpha}(x)&=\left(\frac{\omega_{\zeta}}{2\pi}\right)^{1/4}\exp\left(-\frac{\xi^2}{2}\right)\sum_{n=1}^{\infty}\frac{(\tilde{\alpha}/\sqrt{2})^{n}}{n!}\sqrt{2n}H_{n-1}(\xi), \label{77} \\ \psi_{\alpha}(x)&=\left(\frac{\omega_{\zeta}}{2\pi}\right)^{1/4}\exp\left(-\frac{\xi^2}{2}-\frac{\tilde{\alpha}^2}{2}+\sqrt{2}\tilde{\alpha}\xi\right). \label{78}
	\end{align}
\end{subequations}
Eq.~(\ref{78}) corresponds to the wave function of an un-normalized standard coherent state.

\section{Phase-space representation}

\subsection{Properties of the Wigner function in quantum mechanics}
The Wigner function $W(\vec{r},\vec{p})$ is the cornerstone of quantum mechanics in phase space. It is a quasiprobability distribution defined as \cite{w32,hosw84,Cahill,Kenfack}
\begin{equation}
W(\vec{r},\vec{p})=\frac{1}{\left(2\pi\hbar\right)^n}\int_{-\infty}^{\infty}\exp\left(\frac{i}{\hbar}\,\vec{p}\cdot\vec{r}'\right)\left\langle\vec{r}-\frac{\vec{r}'}{2}\right\vert\rho\left\vert\vec{r}+\frac{\vec{r}'}{2}\right\rangle d\vec{r}',
\label{WM}
\end{equation}
where $\rho$ is the density matrix; $\vec{r}=(r_1,r_2,\dots,r_n)$ and $\vec{p}=(p_1,p_2,\dots,p_n)$ are $n$-dimensional vectors representing the classical phase-space position and momentum values, respectively; and $\vec{r}'=(r'_1,r'_2,\dots,r'_n)$ is the position vector in the integration process. The normalization condition is given by
\begin{equation}
\int_{-\infty}^{\infty}\int_{-\infty}^{\infty}W(\vec{r},\vec{p})d\vec{r}\,d\vec{p}=1.
\end{equation}
It is a real function that can take negative values, in contrast with the probability density of any quantum state. Such {\it negativity} has no physical meaning if it is thought of like a probability distribution. In several works, the negativity in the WF is considered as an indicator of the nonclassicality of a state and being interpreted as a sign of {\it quantumness}. This feature has been tested experimentally \cite{l69,w84,r85,sbrf93,dwm95,bw96,bsm97,brwk99,lszwm03}. However in two-dimensional phase space, for instance, the WF can be used for obtaining the $x$ and $p$ probability distributions 
\begin{equation}
\int_{-\infty}^{\infty}W(x,p)dp=\vert\psi(x)\vert^2, \quad \int_{-\infty}^{\infty}W(x,p)dx=\vert\varphi(p)\vert^2.
\end{equation}

\subsubsection{Wigner function for Landau states in uniaxially strained graphene}

To calculate the Wigner matrix for the Landau state $n$ in the valley $K_D$, we have to take into account that the $n \neq 0$ level presents a four-fold degeneracy due to the pseudospin of the sublattice and valley, while the ground state $n = 0$ has only two-fold degeneracy \cite{Novoselov,Zhang}. We perform our analysis in the $K_D$ valley finding a $2 \times 2$ WM when we substitute the $n$-th eigenstate in Eq.~(\ref{17}) into the Wigner representation \eqref{WM}:
\begin{equation}\label{6}
	W_{n}(\vec{r},\vec{p})=\frac{1}{2^{(1-\delta_{0n})}}W(y,p_y)\left(\begin{array}{c c}
	(1-\delta_{0n})W_{n-1,n-1}(x,p_{x}) & -i\,\lambda(1-\delta_{0n})W_{n-1,n}(x,p_{x}) \\
	i\,\lambda(1-\delta_{0n})W_{n,n-1}(x,p_{x}) & W_{n,n}(x,p_{x})
	\end{array}\right),
\end{equation}
where the components $W_{\alpha,\beta}(x,p_x)$ and $W(y,p_y)$ are given, respectively, by
\begin{subequations}
\begin{align}
		W_{\alpha,\beta}(x,p_x)&=\frac{1}{\pi\hbar}\int_{-\infty}^{\infty}\exp\left(2\frac{i}{\hbar}p_xz_1\right)\psi_{\alpha}(x-z_1)\psi_{\beta}^\ast(x+z_1) dz_1, \label{18} \\
		W(y,p_y)&=\frac{1}{\pi\hbar}\int_{-\infty}^{\infty}\exp\left(2i\left(\frac{p_y}{\hbar}-k\right)z_2\right)dz_2\equiv\delta\left(p_y-k\hbar\right),
\end{align}
\end{subequations}
$\psi_{\alpha}$ and $\psi_{\beta}$ being the wave functions of the quantum harmonic oscillator \eqref{5}.

For computing the function $W_{\alpha,\beta}(x,p_x)$, we define the following quantities
\begin{equation}\label{19}
	\xi=\sqrt{\frac{\omega_{\zeta}}{2}}\left(x+\frac{2k}{\zeta\omega_\zeta}\right), \quad y=\sqrt{\frac{\omega_{\zeta}}{2}}\frac{z_1}{\hbar}, \quad s=\sqrt{\frac{2}{\omega_{\zeta}}}p_x.
\end{equation}
Hence, by substituting Eq.~(\ref{5}) in Eq.~(\ref{18}) and using the definitions (\ref{19}), we get
\begin{equation}\label{38}
	W_{\alpha,\beta}(\chi)=\frac{\exp\left(-\frac{1}{2}|\chi|^2\right)}{\pi}\times\left\{\begin{array}{c c}
		(-1)^\alpha\sqrt{\frac{\alpha!}{\beta!}}\chi^{\beta-\alpha}L_{\alpha}^{\beta-\alpha}\left(|\chi|^2\right), & \text{if } \alpha\leq\beta, \\
		(-1)^\beta\sqrt{\frac{\beta!}{\alpha!}}\chi^{*\alpha-\beta}L_{\beta}^{\alpha-\beta}\left(|\chi|^2\right), & \text{if } \alpha\geq\beta,
	\end{array}\right.
\end{equation}
where the functions $L_n^m(x)$ are the associated Laguerre polynomials and the quantity $\chi = \sqrt{2}(\xi +is)$ is defined. Thus, the components of the $2\times2$ WM turn out to be
\begin{subequations}\label{23}
	\begin{align}
	W_{n-1,n-1}(\chi)&=\frac{1}{\pi}(-1)^{n-1}\exp\left(-\frac{1}{2}|\chi|^2\right)L_{n-1}\left(|\chi|^2\right), \\
	W_{n-1,n}(\chi)&=W_{n,n-1}^\ast(\chi)=\frac{(-1)^{n-1}}{\pi\sqrt{n}}\chi \exp\left(-\frac{1}{2}|\chi|^2\right)L_{n-1}^{1}\left(|\chi|^2\right), \\
	W_{n,n}(\chi)&=\frac{1}{\pi}(-1)^{n}\exp\left(-\frac{1}{2}|\chi|^2\right)L_{n}\left(|\chi|^2\right).
	\end{align}
\end{subequations}
An identical solution for the WM in Eq. \eqref{23} is found in Appendix \ref{A} using the Moyal star product. In these components we can identify the quantity $\mathcal{E}=\tfrac{1}{2}|\chi|^2 = \xi^2+s^2$ or, more explicitly,
\begin{equation}\label{classical}
\mathcal{E}=\frac{2}{\omega_{\zeta}}p_x^2+\frac{\omega_{\zeta}}{2}\left(x+\frac{2k}{\zeta\omega_{\zeta}}\right)^2=\exp\left(2\tau\right)p_x^2+\exp\left(-2\tau\right)\left(x+\frac{2k}{\zeta\omega_{\zeta}}\right)^2,
\end{equation}

\begin{figure}
	\centering
\begin{tabular}{ccc}
(a) \qquad $\epsilon = -20 \%$ \qquad \qquad & (b) \qquad $\epsilon = 0 \%$ \qquad \qquad & (c) \qquad $\epsilon = 20 \%$ \qquad \qquad \\
\includegraphics[trim = 0mm 0mm 0mm 0mm, scale= 0.5, clip]{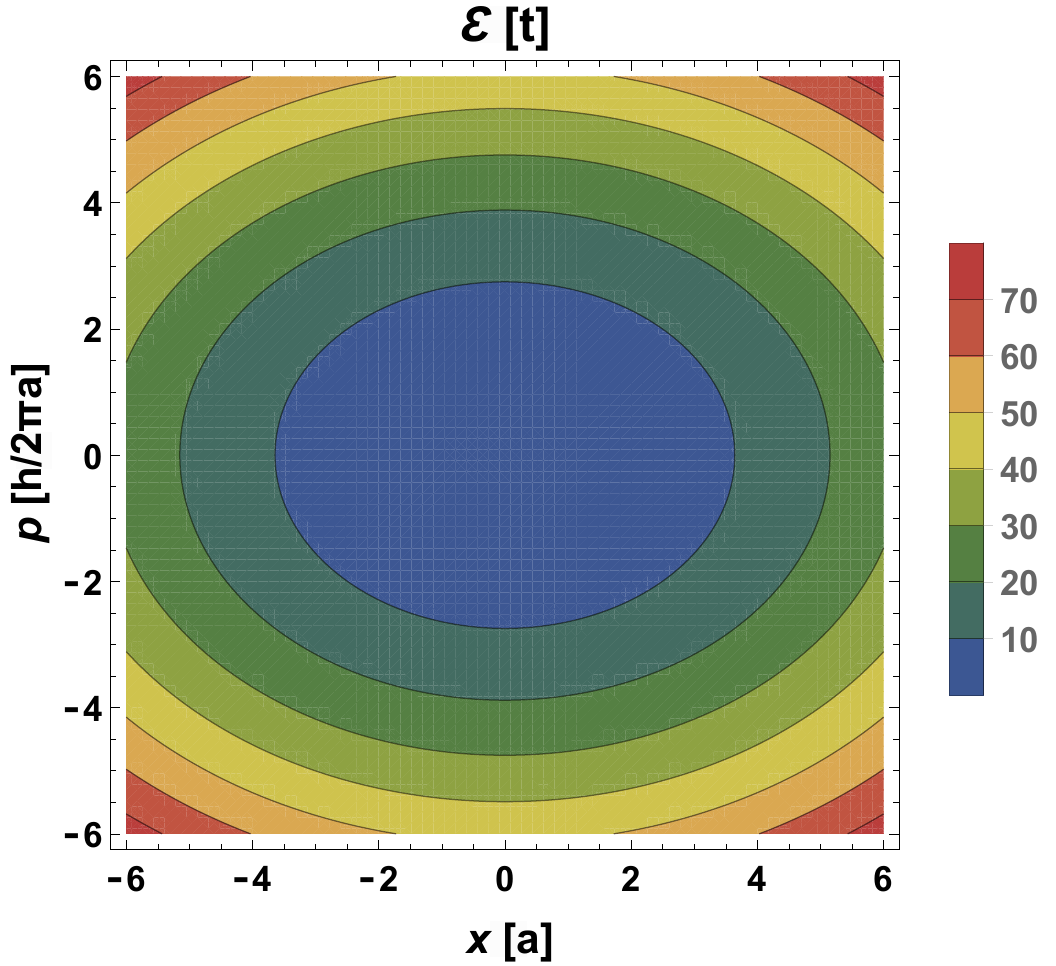} &
\includegraphics[trim = 0mm 0mm 0mm 0mm, scale= 0.5, clip]{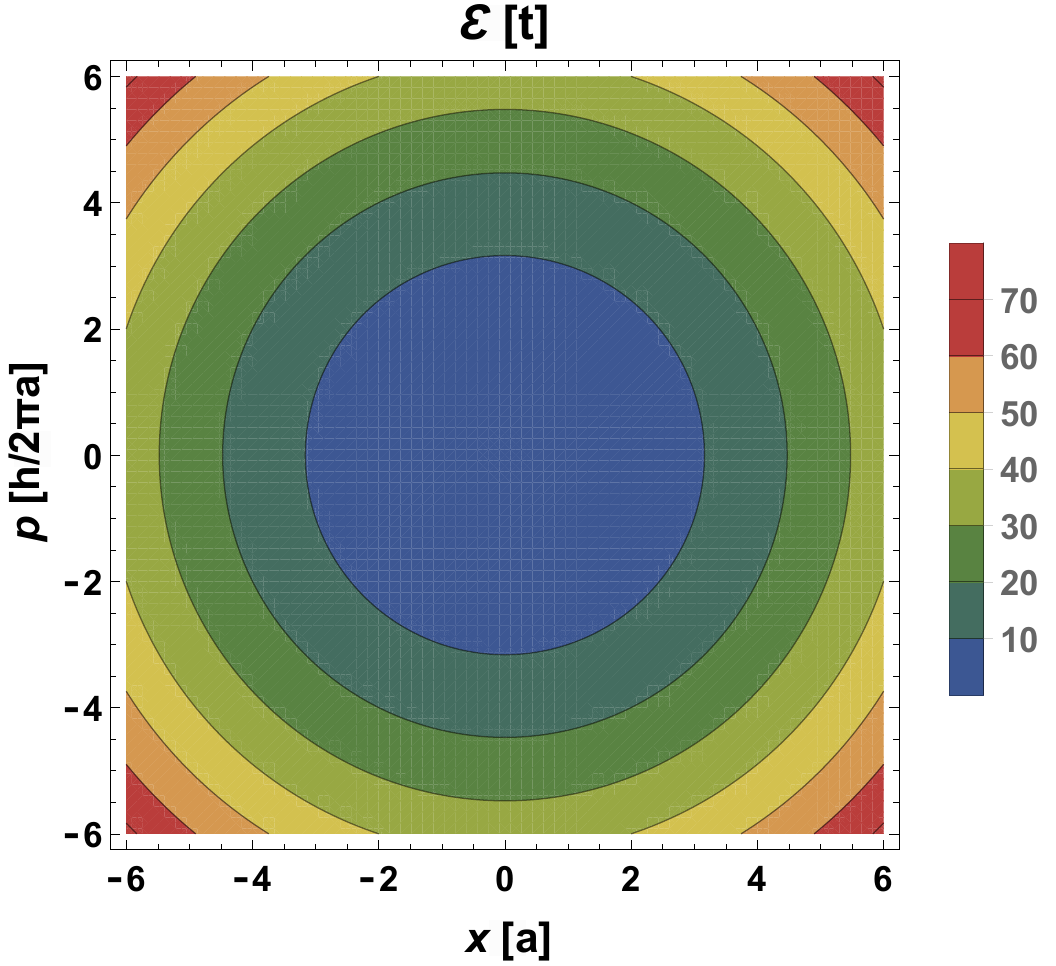} &
\includegraphics[trim = 0mm 0mm 0mm 0mm, scale= 0.5, clip]{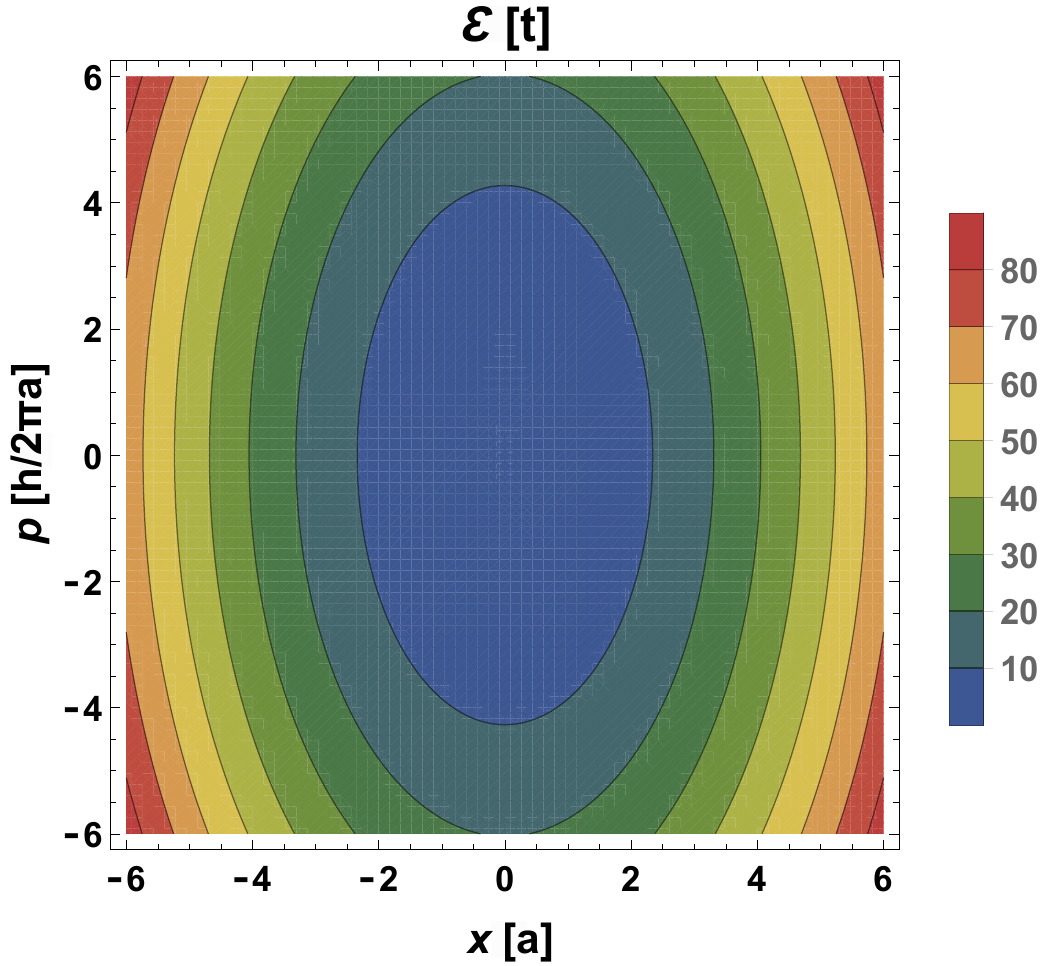}
\end{tabular}
	\caption{Qualitative behavior of the classical energy $\mathcal{E}$ for different values of the tensile strain $\epsilon$ along the zigzag direction. $B = 1$ and $k = 0$.}
	\label{ClassE}
\end{figure}

\begin{figure}[h!]
	\centering
	\begin{minipage}[b]{0.42\textwidth}
		\includegraphics[width=\textwidth]{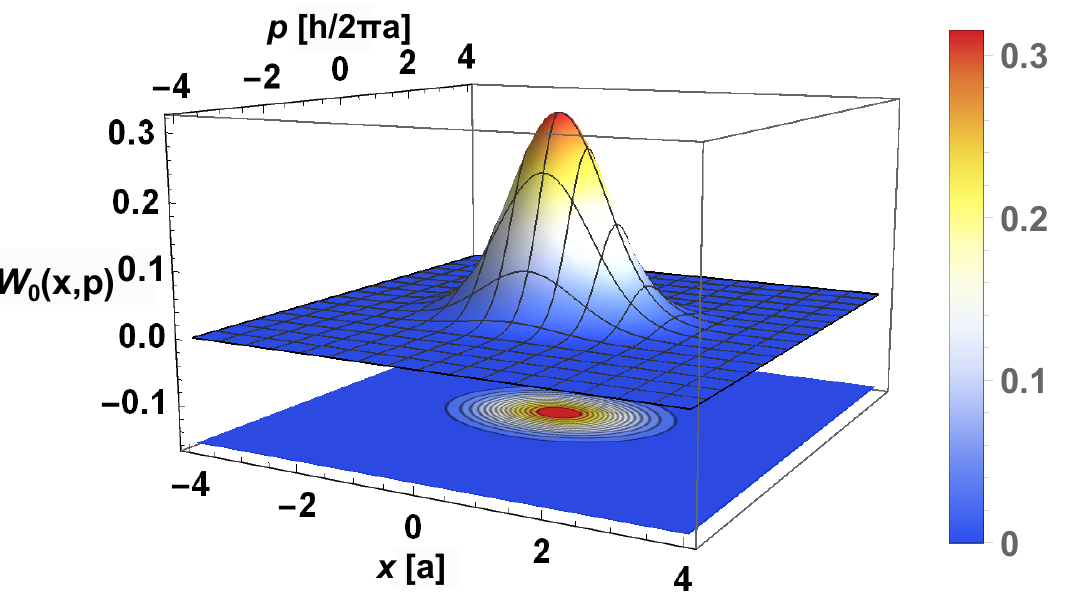}\\
		\centering{\footnotesize (a) $n=0$, $\epsilon=-20 \%$}
		\label{fig:WgStr_0}
	\end{minipage}
	\hspace{1cm}
	\begin{minipage}[b]{0.42\textwidth}
		\includegraphics[width=\textwidth]{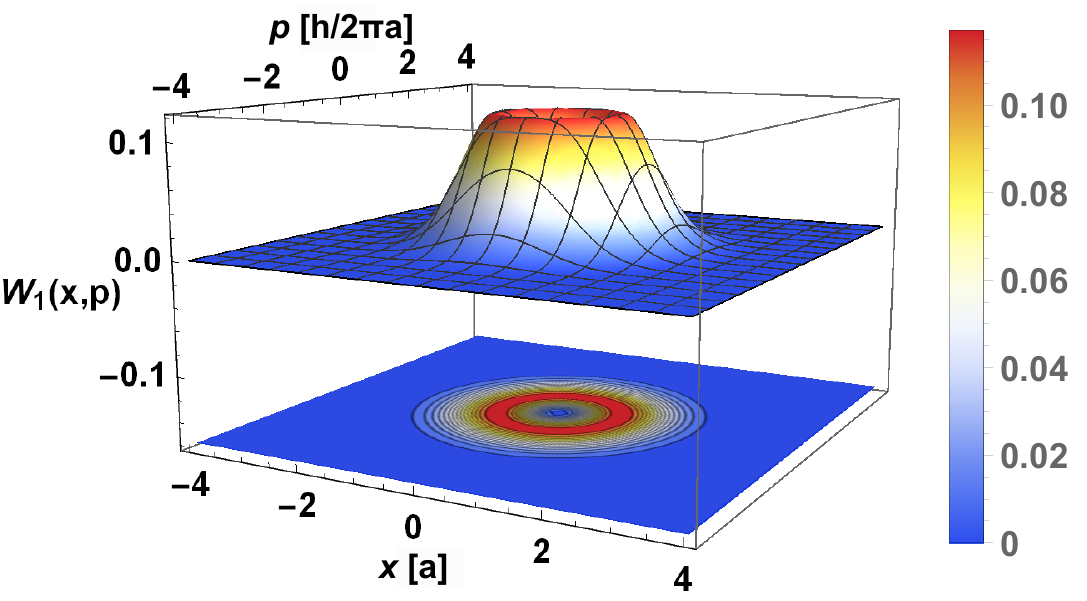}\\
		\centering{\footnotesize (b) $n=1$, $\epsilon=0 \%$}
		\label{fig:WgStr_1}
	\end{minipage}
	\hspace{1cm}
	\begin{minipage}[b]{0.42\textwidth}
		\includegraphics[width=\textwidth]{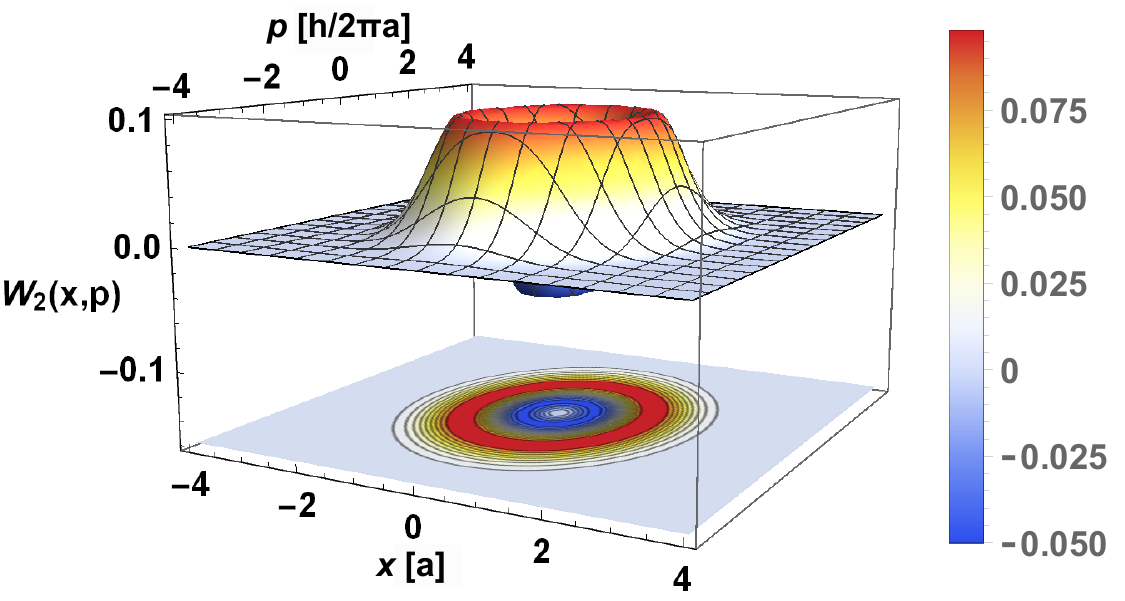}\\
		\centering{\footnotesize (c) $n=2$, $\epsilon= 10 \%$}
		\label{fig:WgStr_2}
	\end{minipage}
	\hspace{1cm}
	\begin{minipage}[b]{0.42\textwidth}
		\includegraphics[width=\textwidth]{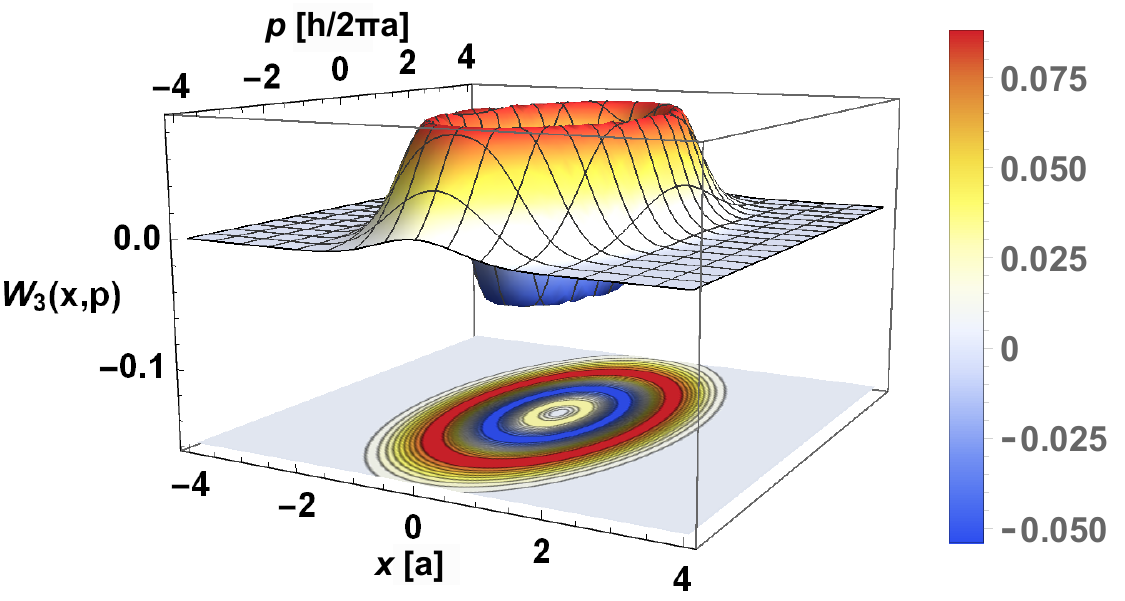}\\
		\centering{\footnotesize (d) $n=3$, $\epsilon= 20 \%$}
		\label{fig:WgStr_3}
	\end{minipage}
	\caption{\label{fig:rhoN}Trace of the Wigner matrix $W_{n}(\vec{r},\vec{p})$ in Eq.~(\ref{6}) for different values of $n$ and $\epsilon$ along the zigzag direction. $B=1$ and $k=0$.}
\end{figure}

\begin{figure}[h!]
	\centering
	\begin{minipage}[b]{0.42\textwidth}
		\includegraphics[width=\textwidth]{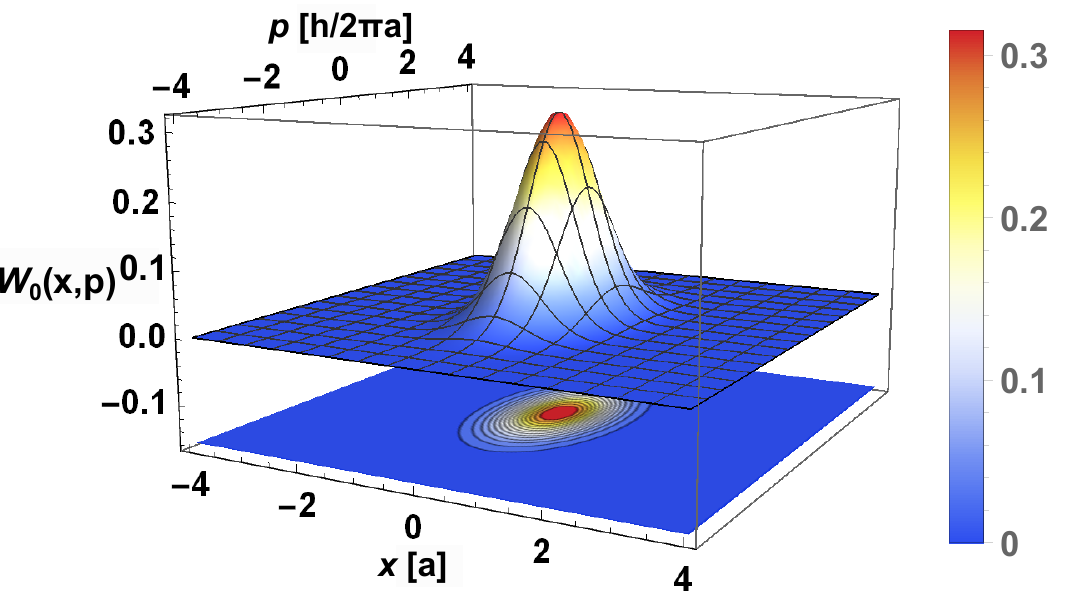}\\
		\centering{\footnotesize (a) $n=0$, $\epsilon=-20 \%$}
		\label{fig:WgStr_0AC}
	\end{minipage}
	\hspace{1cm}
	\begin{minipage}[b]{0.42\textwidth}
		\includegraphics[width=\textwidth]{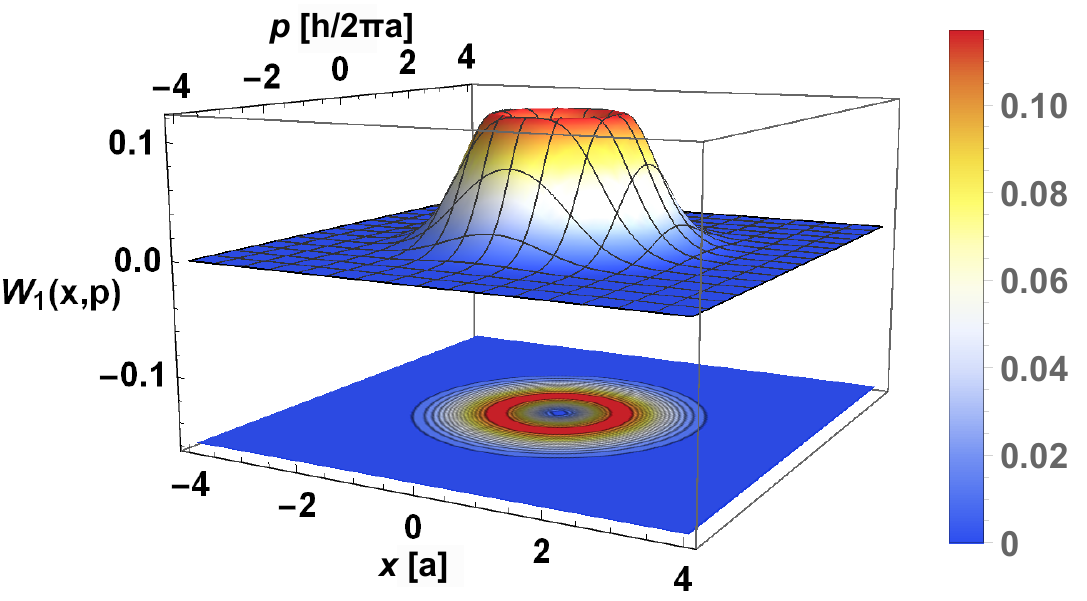}\\
		\centering{\footnotesize (b) $n=1$, $\epsilon=0 \%$}
		\label{fig:WgStr_1AC}
	\end{minipage}
	\hspace{1cm}
	\begin{minipage}[b]{0.42\textwidth}
		\includegraphics[width=\textwidth]{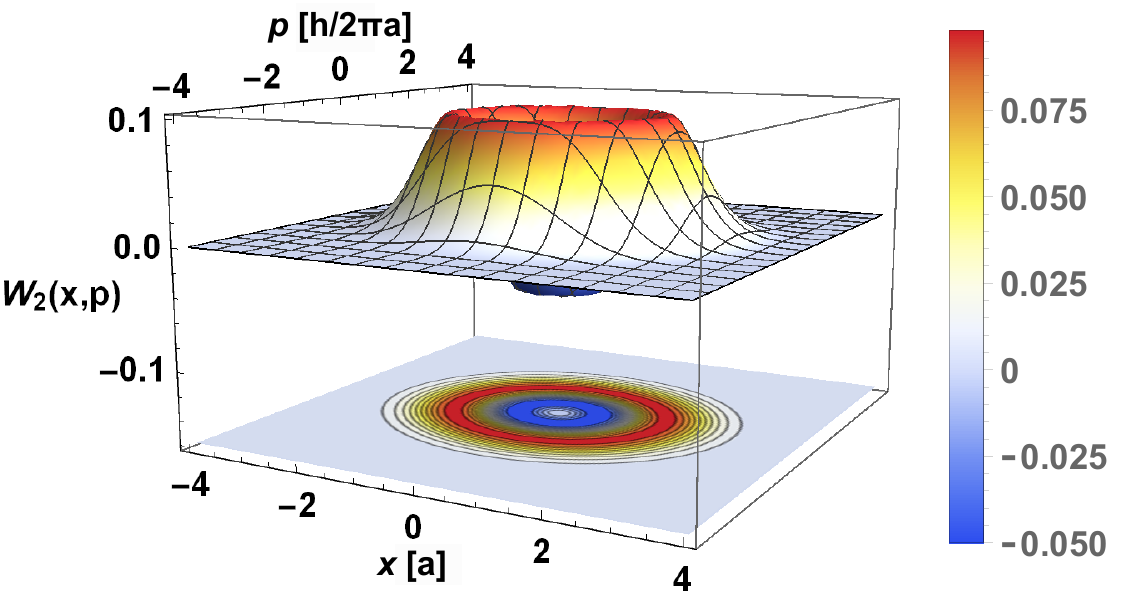}\\
		\centering{\footnotesize (c) $n=2$, $\epsilon=10 \%$}
		\label{fig:WgStr_2AC}
	\end{minipage}
	\hspace{1cm}
	\begin{minipage}[b]{0.42\textwidth}
		\includegraphics[width=\textwidth]{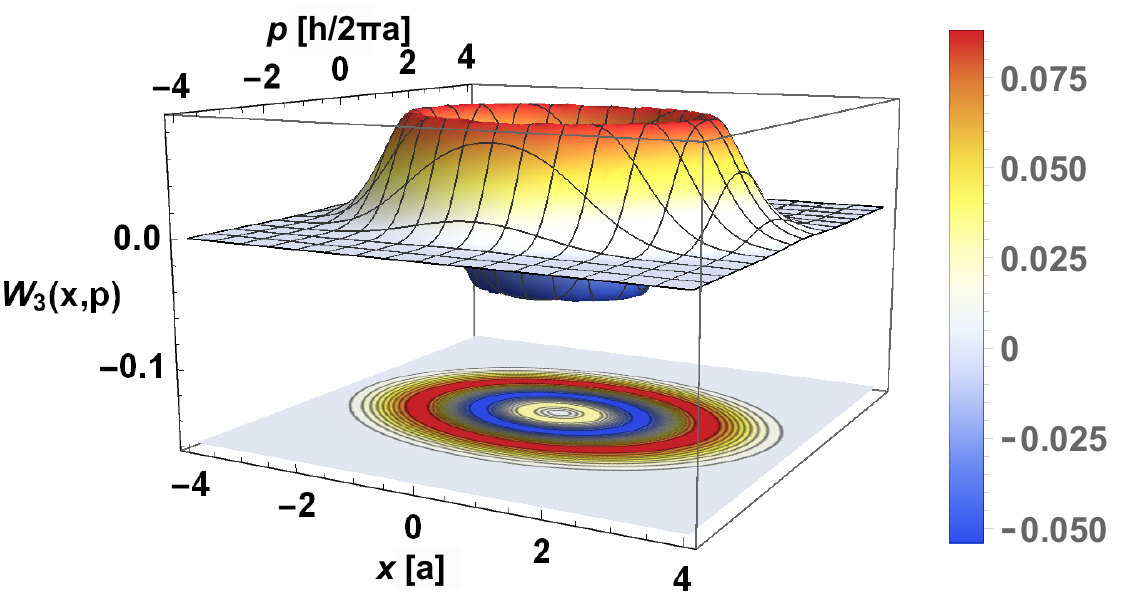}\\
		\centering{\footnotesize (d) $n=3$, $\epsilon=20 \%$}
		\label{fig:WgStr_3AC}
	\end{minipage}
	\caption{\label{fig:rhoN_AC}Trace of the Wigner matrix $W_{n}(\vec{r},\vec{p})$ in Eq.~(\ref{6}) for different values of $n$ and $\epsilon$ along the armchair direction. $B=1$ and $k=0$.}
\end{figure}

\noindent where $\tau=\ln\sqrt{2/\omega_{\zeta}}$. Eq. \eqref{classical} is the classical energy of a particle with linear momentum $p_x$ moving in an oscillator potential the center of which is depending on $k$. As it occurs in classical mechanics, $\mathcal{E}$ is represented by ellipses in phase space. The effect of uniaxial strain is to change the length of the semi-major axis according to the value of $\epsilon$ through the amount $\tau$ in Eq.~(\ref{classical}), as shown in Fig. \ref{ClassE}. This fact is recognizable in the WF for the Landau states of uniaxially strained graphene. 

The elements of $W(\vec{r},\vec{p})$ also have a direct physical interpretation. As seen in wave function \eqref{17}, the electron in sublattice A (B) has a quantum harmonic-like oscillator form $(1- \delta_{n0})\psi_{n-1}$ ($\psi_{n}$). Therefore, $W_{nn}$ and $W_{n-1 n-1}$ are the WFs for the electrons with energy $E_n$ and which belong to the sublattice A and B, respectively. Meanwhile, $W_{n-1,n}$ can be recognized as the mixed WF of an electron in the sublattices A and B. This term can be very important for the case where $e$-$e$ interaction is considered for the description of interference phenomena. However, our analysis is restricted to a single particle model. Therefore, we only address the diagonal terms in the WM. Moreover, due to the fact that one can choose an adequate representation in which the WM is diagonal~\cite{Yuan}, we will just focus on analyzing the trace of matrix $W_{n}(\vec{r},\vec{p})$ in \eqref{6} that is an invariant quantity.

Figures \ref{fig:rhoN} and \ref{fig:rhoN_AC} show that the surface of the Wigner function is distorted differently for deformations along the $\mathcal{Z}$ and $\mathcal{A}$ directions with strain values close to $20 \%$. This evidences that the WF changes following the stress-strain relationship of graphene \cite{Colombo,Cadelano}, where anisotropic and nonlinear elastic behavior is obtained for strains up to 20\%, as shown in Figs. \ref{fig:rhoN}(d) and \ref{fig:rhoN_AC}(d) for the particular case $n =3$, while the isotropic behavior occurs for small strain values up to 10\%. It is important to mention that this same effect of strain on the trace of the WM also emerges for $n = 0$ and other excited Landau states, as shown in Figs. \ref{fig:rhoN}(a), (b), and (c). From the Landau level $n = 2$, we observe negative values in the WF that indicate nonclassicality. This feature usually appears for the WF of single-photon states in quantum cavities \cite{Baune}. The resemblance between the effective Dirac-Weyl model in graphene and the Jaynes-Cummings model established in Refs. \cite{Mota,Ojeda,Jellal,Rusin,Dora,Goldman,Schliemann} shows an important connection between quantum optics systems and graphene, and it can be useful as a key point in the preparation of electron coherent states. For that reason, in the next subsection, we study the phase-space representation of electron coherent states as well as its time evolution from the Heisenberg picture.     

\subsubsection{Wigner function for electron coherent states $\Psi_{\alpha}(x,y)$}

\begin{figure}[h!]
	\centering
	\begin{minipage}[b]{0.45\textwidth}
		\includegraphics[width=\textwidth]{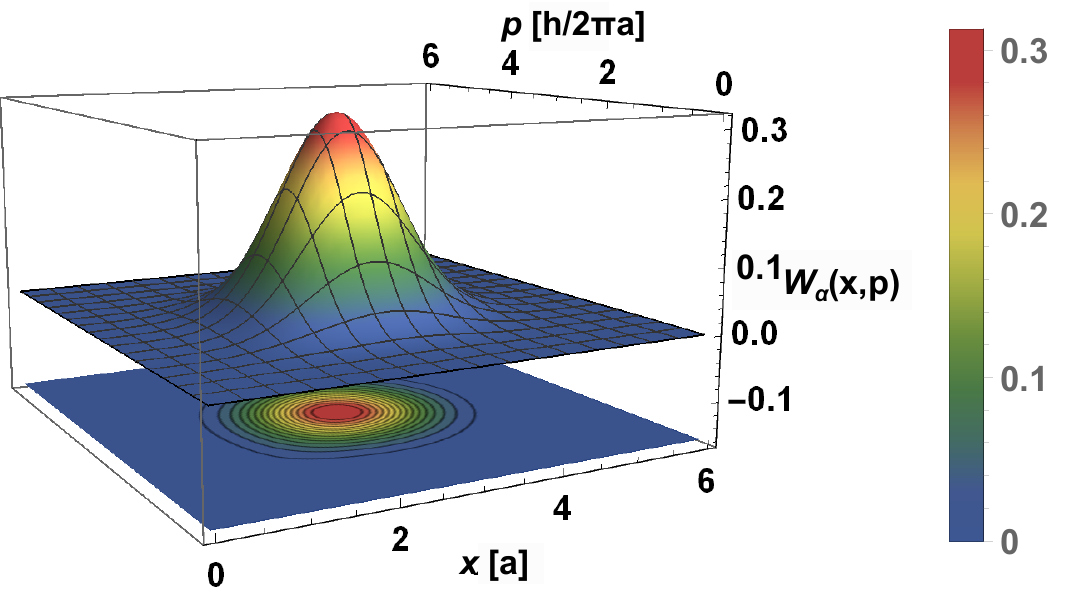}\\
		\centering{\footnotesize (a) $\epsilon_{\mathcal{Z}}=0$}
		\label{fig:WA3}
	\end{minipage}
	\hspace{1cm}
	\begin{minipage}[b]{0.45\textwidth}
		\includegraphics[width=\textwidth]{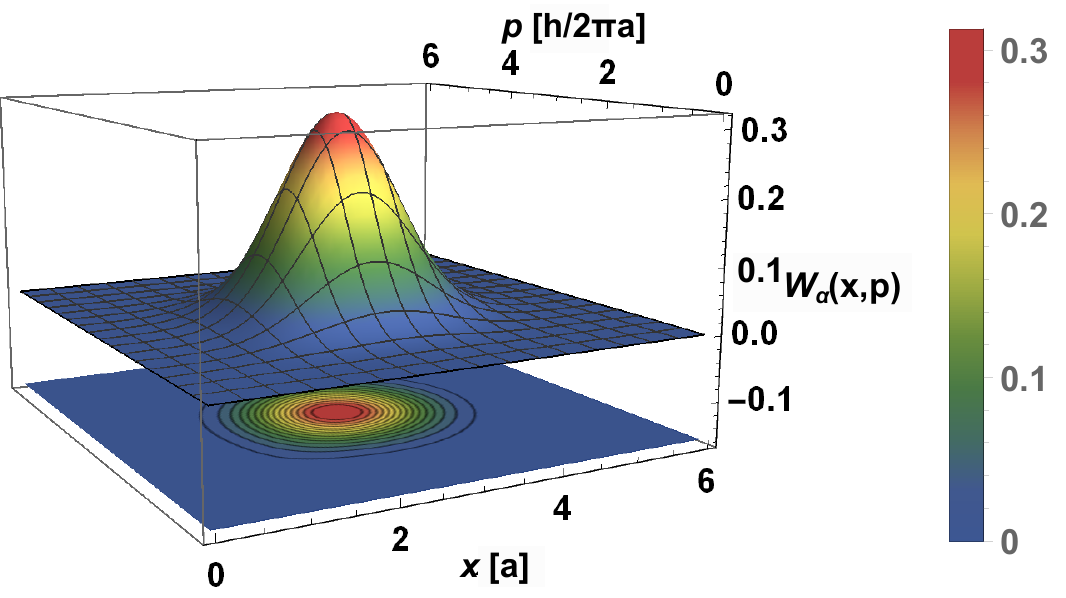}\\
		\centering{\footnotesize (b) $\epsilon_{\mathcal{A}}=0$}
		\label{fig:WA3_AC}
	\end{minipage}
	\begin{minipage}[b]{0.45\textwidth}
		\includegraphics[width=\textwidth]{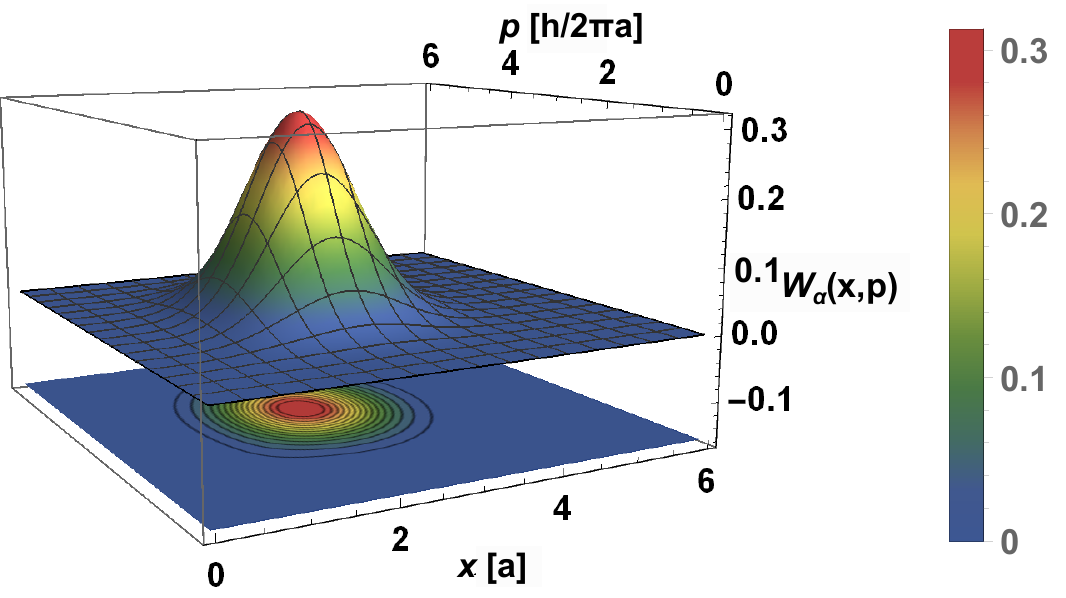}\\
		\centering{\footnotesize (c) $\epsilon_{\mathcal{Z}}=0.1$}
		\label{fig:WA4}
	\end{minipage}
	\hspace{1cm}
	\begin{minipage}[b]{0.45\textwidth}
		\includegraphics[width=\textwidth]{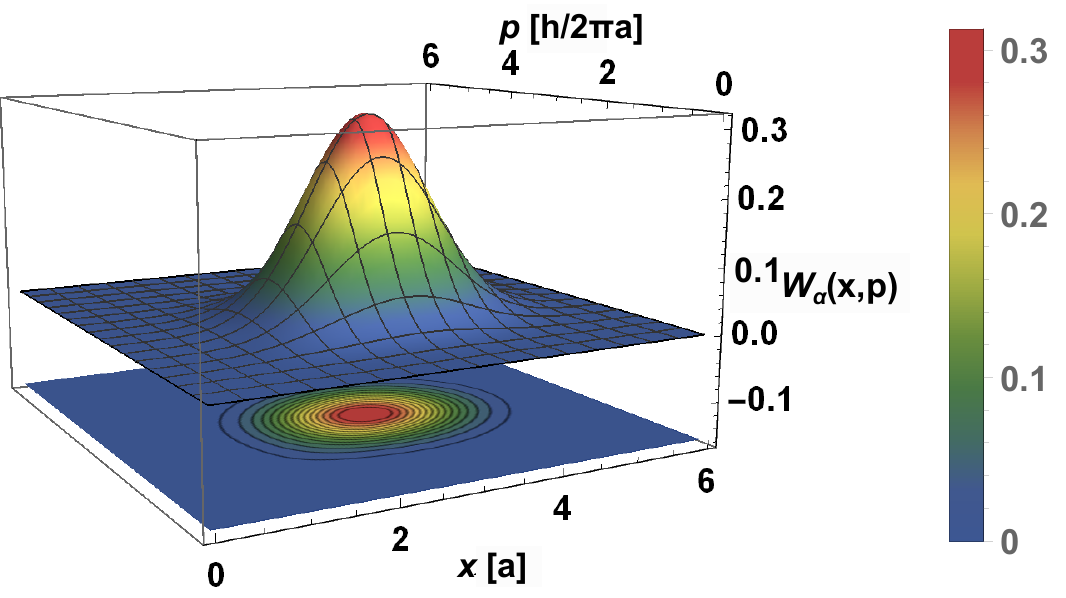}\\
		\centering{\footnotesize (d) $\epsilon_{\mathcal{A}}=0.1$}
		\label{fig:WA4_AC}
	\end{minipage}
	\hspace{1cm}
	\begin{minipage}[b]{0.45\textwidth}
		\includegraphics[width=\textwidth]{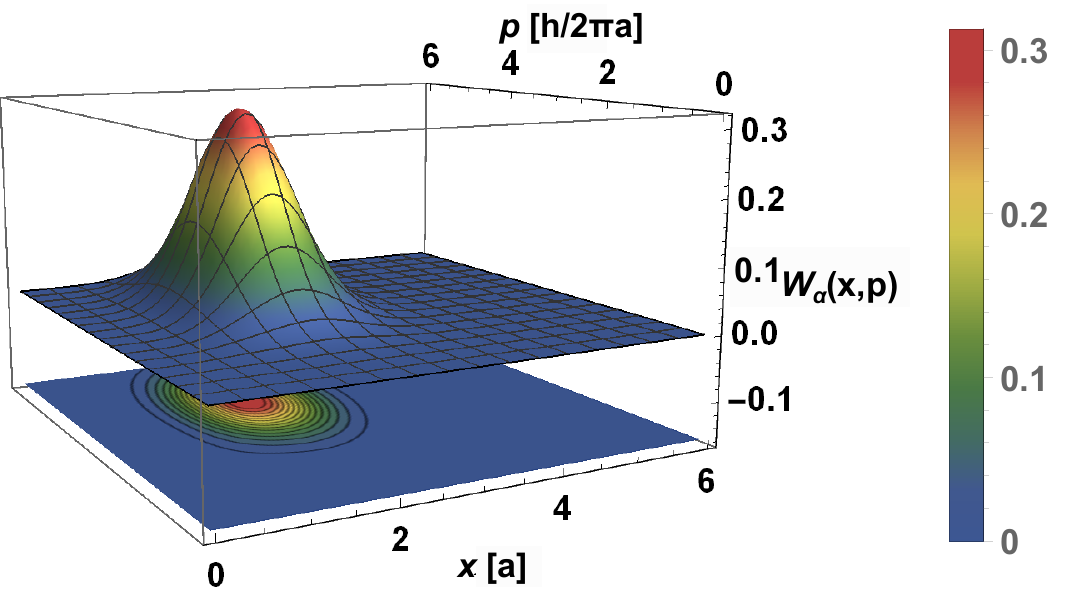}\\
		\centering{\footnotesize (e) $\epsilon_{\mathcal{Z}}=0.2$}
		\label{fig:WA5}
	\end{minipage}
	\hspace{1cm}
	\begin{minipage}[b]{0.45\textwidth}
		\includegraphics[width=\textwidth]{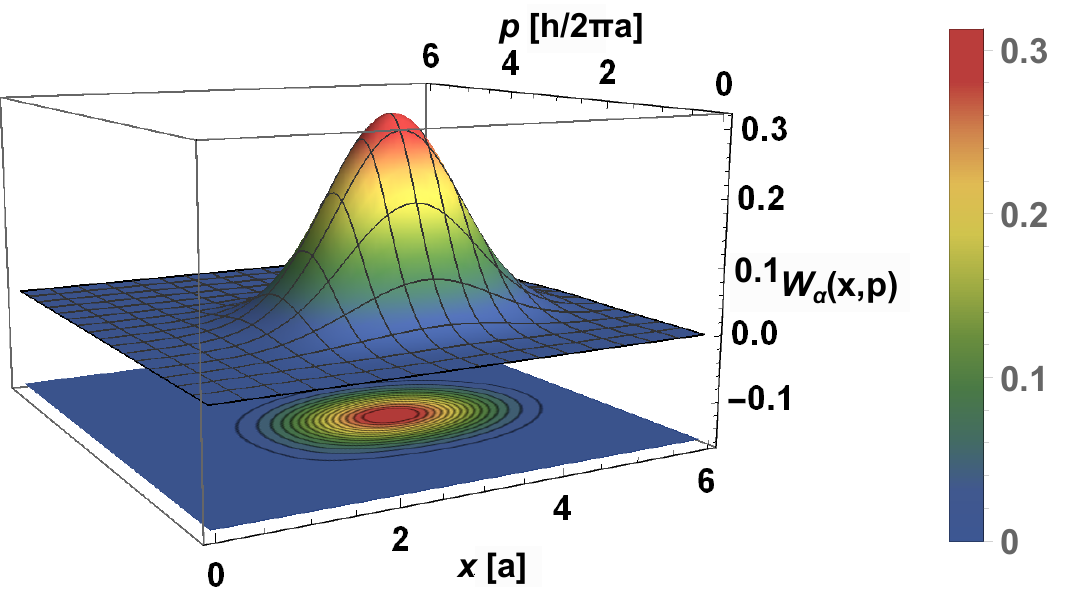}\\
		\centering{\footnotesize (f) $\epsilon_{\mathcal{A}}=0.2$}
		\label{fig:WA5_AC}
	\end{minipage}
	\caption{\label{fig:rhoZZ}Trace of the Wigner matrix $W_{\alpha}(\vec{r},\vec{p})$ from Eq.~(\ref{85}) for deformations in zigzag ((a), (c), and (e)) and armchair directions ((b), (d), and (f)) using the set of values $\alpha=3\exp(i\pi/4)$, $B=1$, $k=\delta=0$, and different values of $\epsilon$.}
\end{figure}

To observe the effect of strain on the WM for electron coherent states, we substitute the components of coherent states \eqref{77} and \eqref{78} into the integral matrix representation \eqref{WM}:
\small
\begin{align}\label{85}
W_{\alpha}(\vec{r},\vec{p})&=\frac{\delta\left(p_y-k\hbar\right)}{\left(2\exp\left(\vert\tilde{\alpha}\vert^2\right)-1\right)}\left(\begin{array}{c c}
W_{11}(x,p_x) & -i\,\lambda W_{12}(x,p_x) \\
i\,\lambda W_{21}(x,p_x) & W_{22}(x,p_x)
\end{array}\right),
\end{align}
\normalsize
where, by using properly Eq.~(\ref{38}), the matrix components are given by
\small
\begin{subequations}\label{components}
\begin{align}
W_{11}(\chi)& = -\frac{\textrm{e}^{-\frac{1}{2}|\chi|^2}}{\pi}\sum_{n=1}^{\infty}\left[\frac{(-|\tilde{\alpha}|^2)^n}{n!}L_{n-1}\left(|\chi|^2\right) +2\sum_{m=n+1}^{\infty}\frac{(-1)^{n}}{m!}\Re\{\tilde{\alpha}^n\tilde{\alpha}^{\ast m}\chi^{m-n}\}\sqrt{\frac{m}{n}}L_{n-1}^{m-n}\left(|\chi|^2\right)\right], \\
W_{12}(\chi)&=W_{21}^{\ast}(\chi)=\frac{1}{\pi}\exp\left(-\frac{1}{2}|\chi|^2 + \tilde{\alpha}^*\chi\right)\sum_{n=1}^{\infty}\frac{\tilde{\alpha}^{n}}{n!}\sqrt{n} \left(\chi^* - \tilde{\alpha}^*\right)^{n-1}, \\
W_{22}(\chi)&=\frac{1}{\pi}\exp\left(\frac{1}{2}|\chi|^2-|\chi - \tilde{\alpha}|^2\right), \label{44c}
\end{align}
\end{subequations}
\normalsize
where $\Re(z)$ and $\Im(z)$ denote the real and imaginary part of a complex number $z$, respectively. Expression~(\ref{44c}) corresponds to the scalar WF of un-normalized coherent states as expected. The off-diagonal components indicate a sign of interference by the superposition of the electron wave function from the sublattices A and B. Fig.~\ref{fig:rhoZZ} shows the trace of $W_{\alpha}(\vec{r},\vec{p})$ for strains applied in the $\mathcal{Z}$ and $\mathcal{A}$ direction, respectively. We observe that the trace of the WM also behaves as the stress-strain relationship of graphene. Since the compression of LLs favors the realization of electron coherent states, we only consider positive strain values for the WM. By comparing deformations along the $\mathcal{Z}$ and $\mathcal{A}$ direction, the shape of the $W_{\alpha}(\vec{r},\vec{p})$ is different in the nonlinear elastic regime. We can observe for $\epsilon < 10 \%$ that the WF has an identical response under the application of uniaxial strain in $\mathcal{Z}$ and $\mathcal{A}$ directions but with a difference of $90^{\textrm{o}}$. Thus, the strain affects the uncertainty relation, such as distorting the classical energy orbits, (see Fig. \ref{ClassE}). Further, the coherent states keep their minimum values of uncertainty in position and linear momentum \cite{Knight}. Herein, the positive strains along the $\mathcal{Z}$ ($\mathcal{A}$) direction increase (decrease) the uncertainty of $p_x$, causing simultaneously a decrease (increase) in the uncertainty of $x$, according to the amount $\tau$ in Eq.~(\ref{classical}). Therefore, the action of strain on electron coherent states in graphene is similar to the effect found in the squeezing of light coherent states in quantum optics \cite{Knight}.  

On the other hand, when we change the parameter $\alpha$ it is possible to identify two main effects on the WF. First, the increasing of $|\alpha|^2$ shifts the maximum of the WF far away from the phase-space origin. This is identical to the displacement observed in light coherent states, where $|\alpha|^2$ also corresponds to the average photon number \cite{Knight}. Therefore, the quantity $|\alpha|^2$ is proportional to the distance of the maximum of the WF for an electron coherent state. Second, we also observe that the phase of $\alpha$ causes rotations of the WF around the phase-space origin. However, we will show that the time evolution of electron states is very different from those of quantum optics. 

\subsubsection{Time evolution of the Wigner matrix for electron coherent states}
Now, we investigate the time evolution of the coherent states in phase space by applying the time evolution unitary operator $U(t)=\exp\left(-iHt/\hbar\right)$ on the states $\Psi_{\alpha}(x,y)$ in Eq. \eqref{CE}:
\begin{equation}\label{evolutionPsi}
\Psi_{\alpha}(x,y,t)=U(t)\Psi_{\alpha}(x,y)=\frac{1}{\sqrt{2\exp\left(\vert\tilde{\alpha}\vert^2\right)-1}}\left[\Psi_{0}(x,y)+\sum_{n=1}^{\infty}\frac{\sqrt{2}\,\tilde{\alpha}^n}{\sqrt{n!}}\exp\left(-iE_{n}t\right)\Psi_{n}(x,y)\right],
\end{equation}
where $E_{n}$ are the LLs \eqref{LLs}. 

By substituting Eq.~(\ref{evolutionPsi}) in Eq.~(\ref{WM}), the expressions in~(\ref{components}) are recalculated and written, respectively, as
\small
\begin{subequations}\label{time}
	\begin{align}
	& W_{11}(\chi,t) = -\frac{e^{-\frac{1}{2}|\chi|^2}}{\pi}\sum_{n=1}^{\infty}\left[\frac{(-|\tilde{\alpha}|^2)^n}{n!}L_{n-1}\left(|\chi|^2\right)  +2\sum_{m=n+1}^{\infty}\frac{(-1)^{n}}{m!}\chi^{m-n}A_{nm}(t)\sqrt{\frac{m}{n}}L_{n-1}^{m-n}\left(|\chi|^2\right)\right], \\
	&\nonumber W_{12}(\chi,t)=W_{21}^{\ast}(\chi,t)=\frac{e^{-\frac{1}{2}|\chi|^2}}{\pi}\sum_{n=0}^{\infty}\left[\frac{\tilde{\alpha}\vert\tilde{\alpha}\vert^{2n}}{n!\sqrt{n+1}}(-1)^{n}e^{i(E_{n}-E_{n+1})t}L_{n}\left(|\chi|^2\right) \right. \\
& \qquad\qquad\qquad\qquad\qquad \left. +2\sum_{m=n+1}^{\infty}\frac{(-1)^{n}}{m!\sqrt{n+1}}\chi^{m-n}A_{n+1 m}(t)L_{n-1}^{m-n}\left(|\chi|^2\right)\right], \\
	& W_{22}(\chi,t)=\frac{e^{-\frac{1}{2}|\chi|^2}}{\pi}\left[J_{0}\left(2i|\tilde{\alpha\chi}|\right)e^{-|\tilde{\alpha}|^2} +2\sum_{n=0}^{\infty}\sum_{m=n+1}^{\infty}\frac{(-1)^{n}}{m!}\chi^{m-n}A_{nm}(t)L_{n}^{m-n}\left(|\chi|^2\right)\right],
	\end{align}
\end{subequations}
\normalsize
\begin{figure}[h!]
	\centering
	\begin{minipage}[b]{0.45\textwidth}
		\includegraphics[width=\textwidth]{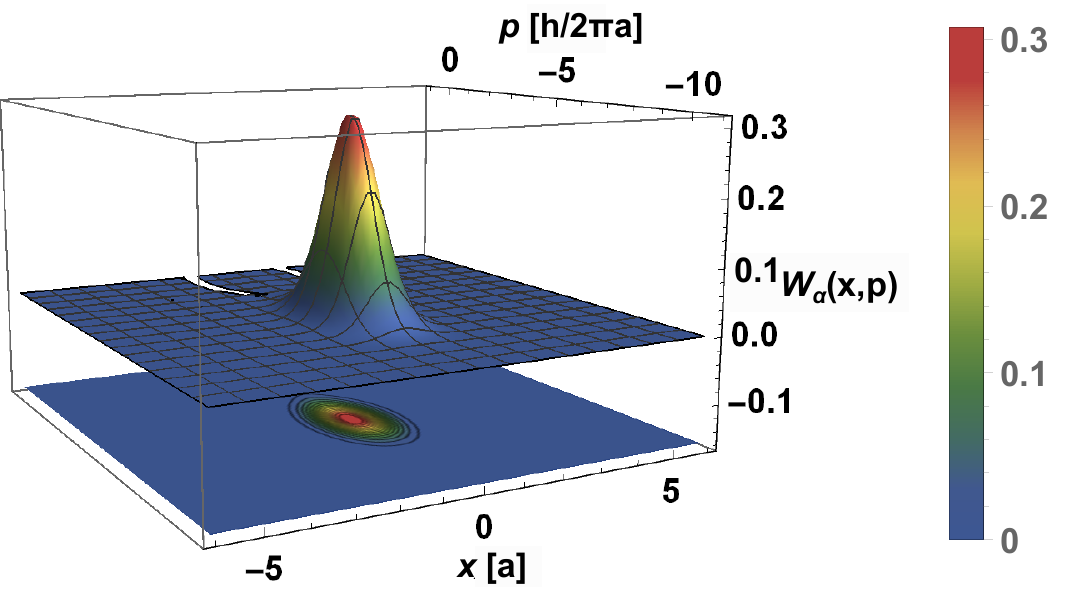}\\
		\centering{\footnotesize (a) $t'=0$}
		\label{fig:Wtime00}
	\end{minipage}
	\hspace{1cm}
	\begin{minipage}[b]{0.45\textwidth}
		\includegraphics[width=\textwidth]{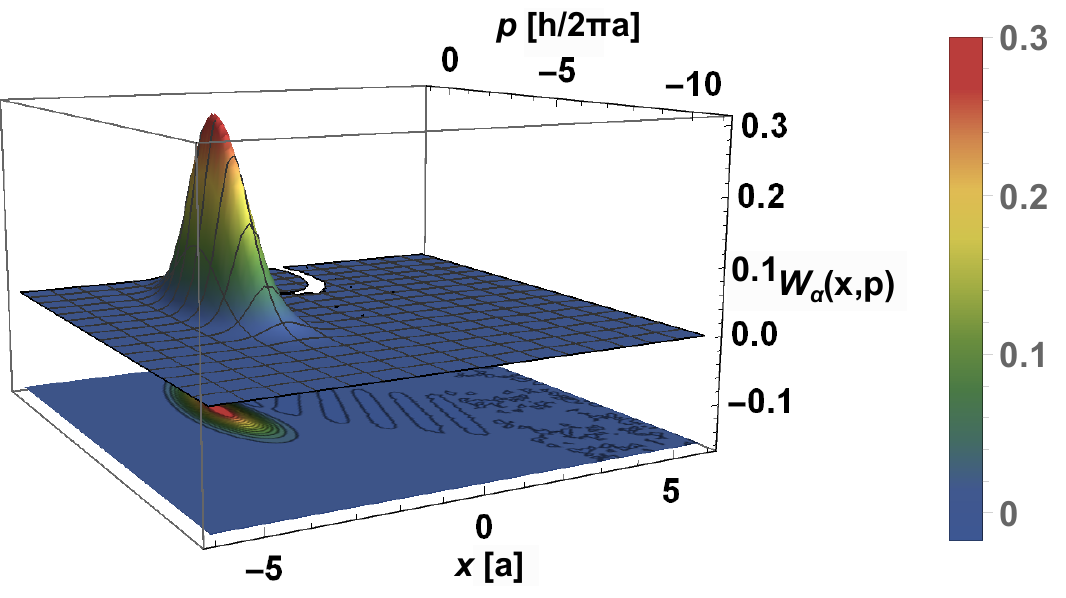}\\
		\centering{\footnotesize (b) $t'=5$}
		\label{fig:Wtime01}
	\end{minipage}
	\hspace{1cm}
	\begin{minipage}[b]{0.45\textwidth}
		\includegraphics[width=\textwidth]{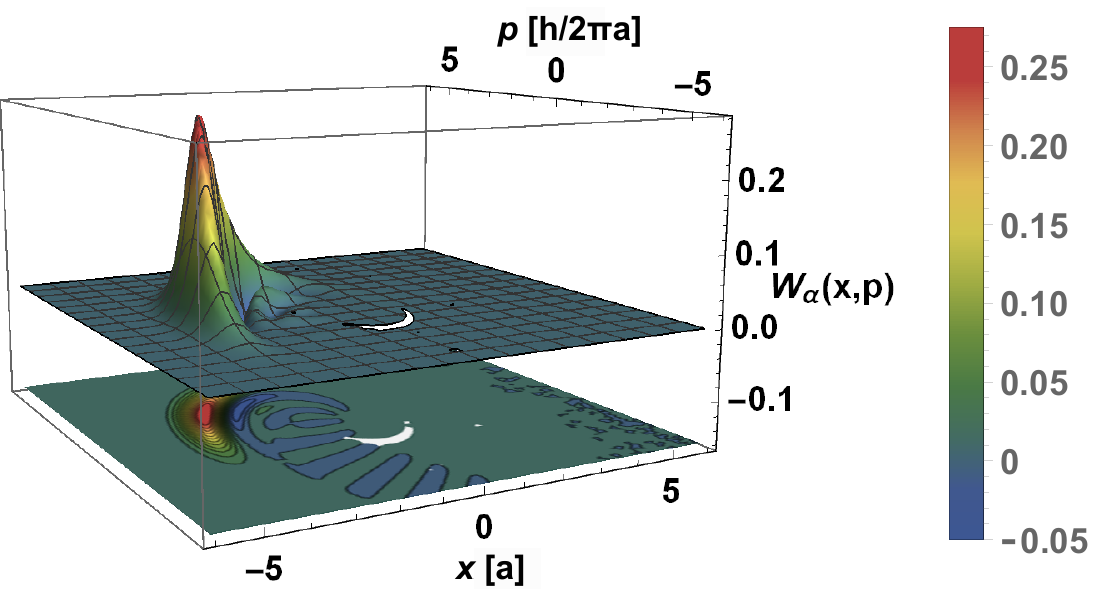}\\
		\centering{\footnotesize (c) $t'=10$}
		\label{fig:Wtime02}
	\end{minipage}
	\hspace{1cm}
	\begin{minipage}[b]{0.45\textwidth}
		\includegraphics[width=\textwidth]{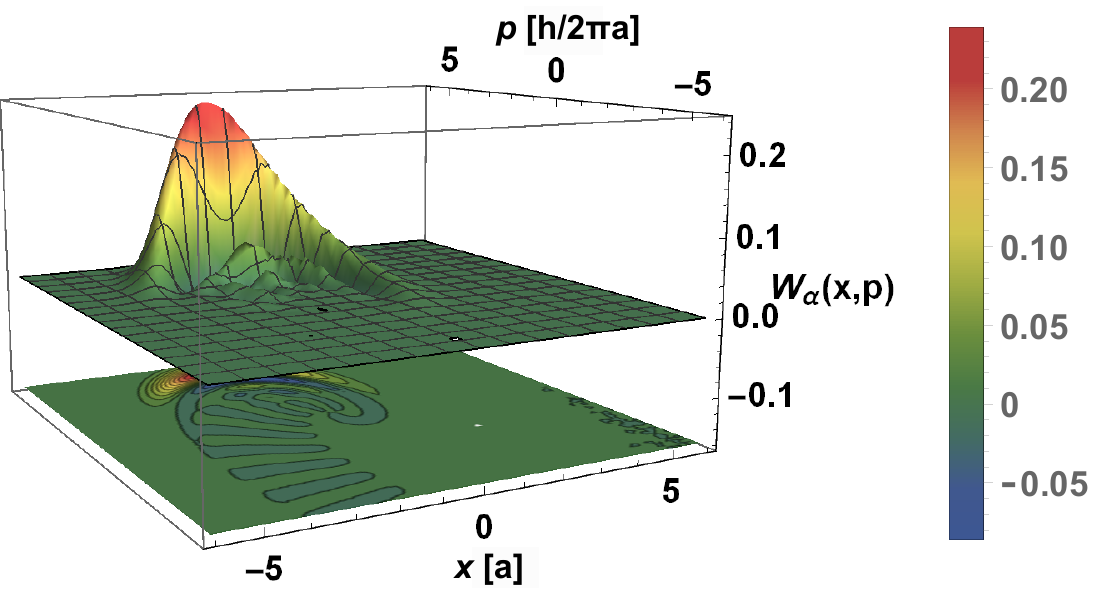}\\
		\centering{\footnotesize (d) $t'=15$}
		\label{fig:Wtime03}
	\end{minipage}
	\hspace{1cm}
	\begin{minipage}[b]{0.45\textwidth}
		\includegraphics[width=\textwidth]{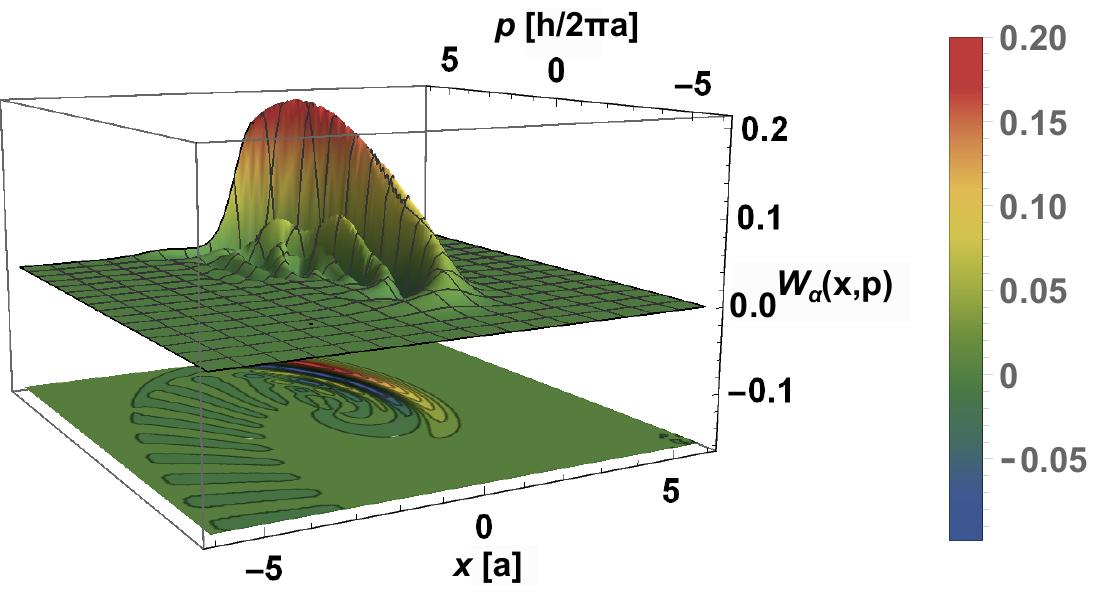}\\
		\centering{\footnotesize (e) $t'=20$}
		\label{fig:Wtime04}
	\end{minipage}
	\hspace{1cm}
	\begin{minipage}[b]{0.45\textwidth}
		\includegraphics[width=\textwidth]{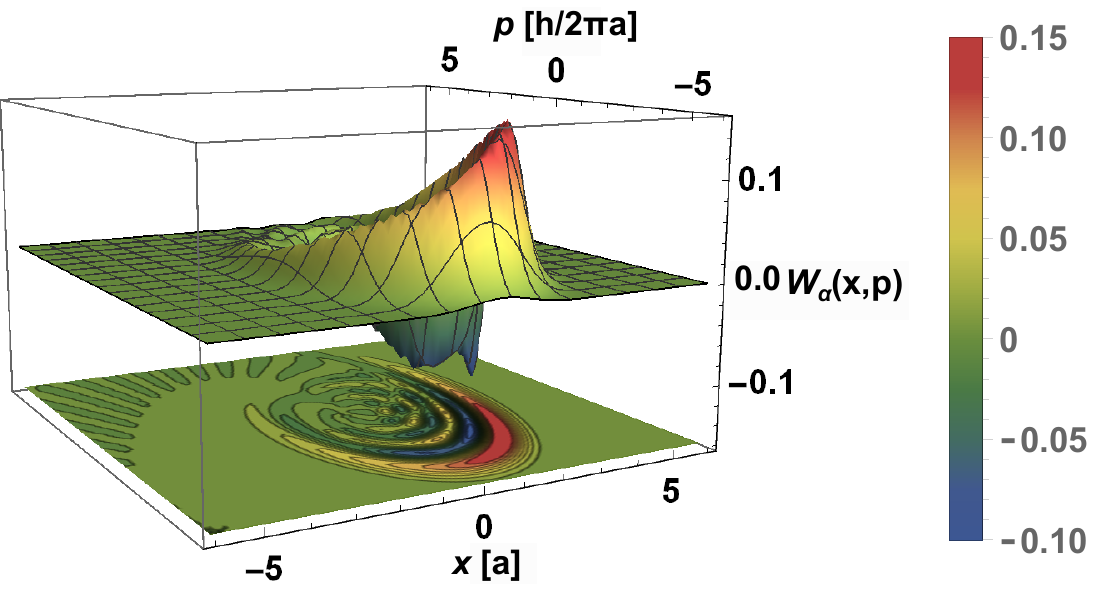}\\
		\centering{\footnotesize (f) $t'=30$}
		\label{fig:Wtime05}
	\end{minipage}
	\hspace{1cm}
\begin{minipage}[b]{0.45\textwidth}
	\includegraphics[width=\textwidth]{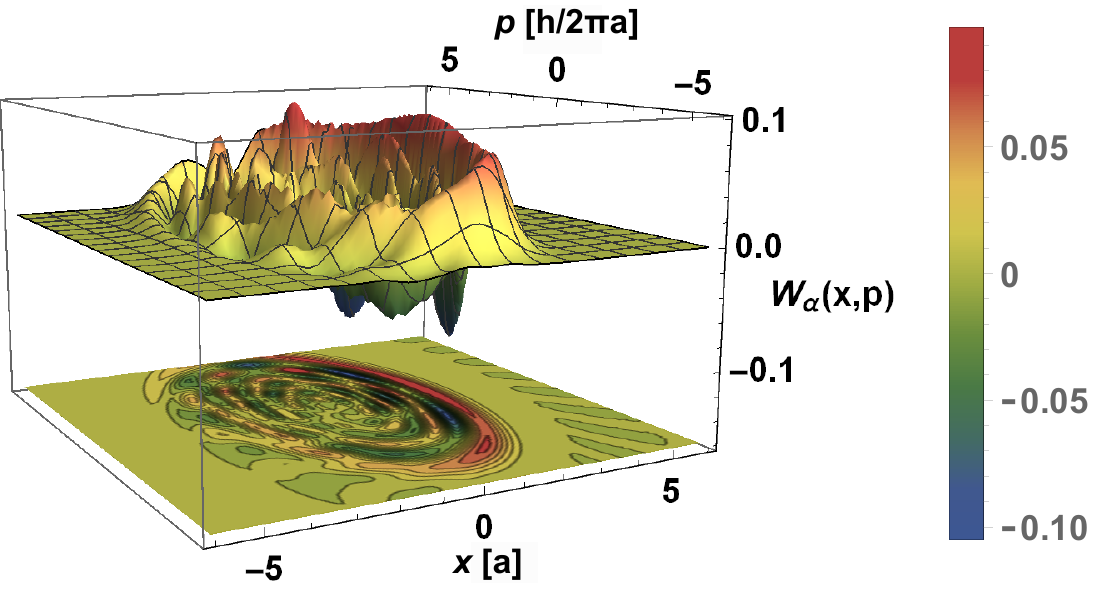}\\
	\centering{\footnotesize (g) $t'=60$}
	\label{fig:Wtime06}
\end{minipage}
\hspace{1cm}
\begin{minipage}[b]{0.45\textwidth}
	\includegraphics[width=\textwidth]{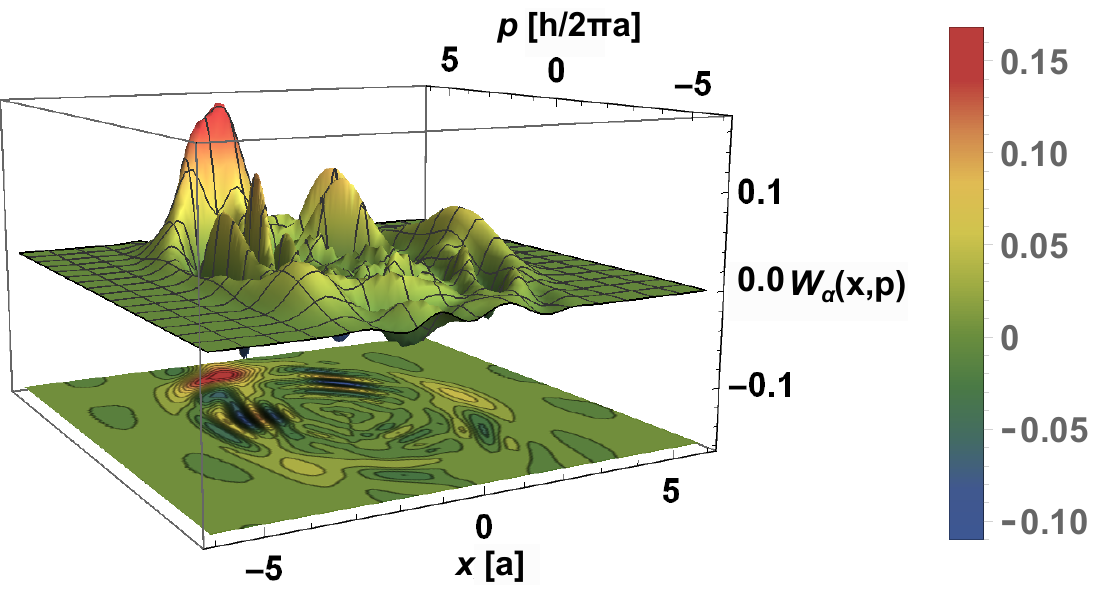}\\
	\centering{\footnotesize (h) $t'=540$}
	\label{fig:Wtime07}
\end{minipage}
	\caption{\label{fig:Wignertime}Time evolution of the trace of the WM $W_{\alpha}(\vec{r},\vec{p},t)$ with $\alpha=3\exp(-i\pi/2)$ and $\epsilon=0.2$ in the $\mathcal{Z}$ direction by using the components in Eq.~(\ref{time}). $B=1$, $k=\delta=0$, and $t'\equiv t/(v'_{\rm F}\sqrt{2e\hbar B})$.}
\end{figure}
\noindent where the identity
	\begin{equation}
J_{\nu}(2\sqrt{xz})\exp\left(z\right)\left(xz\right)^{-\nu/2}=\sum_{n=0}^{\infty}\frac{z^{n}}{\Gamma(n+\nu+1)}L_{n}^{\nu}(x), \quad \nu>-1,
	\end{equation}
is implemented and the temporal function $A_{nm}(t) = \Re(\tilde{\alpha}^n\tilde{\alpha}^{\ast m}e^{i(E_{m}-E_{n})t})$ has been defined. Setting $t = 0$, we recover the expressions \eqref{components}. The $A_{nm}(t)$ indicates that the trace of the WM changes its shape when the time increases due mainly to the fact that $E_{n} \propto \sqrt{n}$. Since positive deformations tend to compress the LL spectra, the time scale $[0,T]$ ($T$ denotes the rotation period) is also contracted as $[0,\sqrt{ab}T]$, while in light coherent states such a contraction does not exist because $E_{n} \propto n$ and the eigenvalue $\alpha$ evolves as $\alpha(t)=\alpha\exp(i\omega t)$. For light coherent states, the variation of the phase of $\tilde{\alpha}$ causes only rotation of the corresponding WF around the phase-space origin without changing its shape \cite{Knight}. However, the LL dependency $\sqrt{n}$ not only produces rotation of the WM when the time increases but also it generates oscillations, as shown in Fig. \ref{fig:Wignertime}. The linear feature in the effective Dirac-Weyl Hamiltonian causes these oscillations in the WF, while in conventional semiconductors it is absent by the quadratic dispersion relation of electrons. As in the Jaynes-Cummings model and Rydberg atoms for a Gaussian wave packet, collapses and revivals begin to manifest during the time evolution of the WF of electron coherent states~\cite{krueckl09}. We can observe that in the interval $0\leq t\leq30$ (Figs.~\ref{fig:Wignertime}(a)-(f)) the maximum value of the WF remains in the time evolution, changing its numerical value and turning around the origin, while negative values emerge through oscillations. This fact represents an important difference between the WFs of electron coherent states in graphene and the light ones in quantum optics. In the former case, we observe that electron coherent states exhibit quantum behavior, while in the latter the WF is always positive in any point of phase space. Nevertheless, the maximum value of the WF for both electron and photons describes closed orbits in phase space. In the present case, this absolute maximum disappears for some values of $t$, as shown in Fig.~\ref{fig:Wignertime}(g). However, it is recovered again after a period (see Fig.~\ref{fig:Wignertime}(h)). Such an effect is comparable to the collapses and revivals of electron wave packets in graphene~\cite{krueckl09}. Furthermore, the time evolution of the WM trace recalls in a sense that of a particle in a Morse potential~\cite{Jalal}. 

Note that, in contrast to that discussed in~\cite{krueckl09}, we have implemented a propagation algorithm that has been applied to a restricted set of problems, for instance~\cite{Rusin,Schliemann,schliemann082}, through the lack of known analytical solutions of the Dirac equation. Hence, due to the uniformity of the magnetic field, we are able to obtain the time evolution of the Wigner matrix by using the integral representation, allowing us to observe that both schemes show revivals like in the time evolution whether in the wave packet or in the WF. Nevertheless, for inhomogeneous magnetic fields and electrostatic potentials, the use of the propagator from Chebyshev polynomials could be very suitable.

\section{Conclusions and final remarks}

We studied the electron dynamics in uniaxially strained graphene in the presence of a uniform and perpendicular magnetic field to the layer from a phase-space representation perspective. The Wigner matrix is a powerful theoretical tool that has been very useful for investigating quantum optics systems and quantum information, and so far has been scarcely used for condensed matter. Since mechanical deformation modifies the atomic positions in the lattice and distorts the reciprocal space, Wigner matrix representation allows evidence of the effect of strain on the electron dynamics simultaneously in both spaces. Moreover, it indicates which electronic states present a nonclassical behavior. This is of relevant importance for the realization of entanglement. By using a tight-binding approach to nearest neighbors, it was possible to derive an effective Dirac-Weyl model, where we obtained Landau-level spectra and their corresponding wave functions. This effective model has a direct connection with other methods frequently used for studying strained materials such as the geometrical approach \cite{Betancur}, where we find an analytical expression of the geometrical parameters $a$ and $b$ as a function of the strain tensor components, as well as connection with the supersymmetric potential model \cite{Concha}. Employing a generalized annihilation operator, we can build the electron coherent states from Landau ones and demonstrate that stretching along the zigzag direction favors the observation of electron coherent states. We have found that the trace of the Wigner matrix for Landau and coherent states has a different response when we apply uniaxial strain along with the zigzag and armchair directions. Also, the time evolution of the Wigner function evidences distinctive features between electronic and light quantum systems. We observe that the maximum value of this function describes closed orbits but, as time increases, the Wigner function shows negative values. This fact contrasts with light coherent states, where positive values of the Wigner function are observed in the whole phase space. 

We think that our findings might help to establish protocols for the preparation of electron coherent states in the laboratory. Moreover, these results suggest the possibility of a feedback between electron and light quantum optics due to the analogy of approaches in condensed matter and quantum optics. Also, it is worth remarking that the phase-space representation presented in this paper has an advantage over canonical quantization, since it has allowed us to investigate the time evolution of electronic coherent states, which is difficult to perform using the notion of the wave function because the energy spectrum (\ref{LLs}) is not equidistant. Also, the so-called Husimi function~\cite{h40} is another tool defined in phase space and used to describe electron dynamics~\cite{datseris19,carmesin19}, so that its implementation in this system could also provide novel information for the condensed matter and quantum optics analogies.

\vspace*{3mm}
\section*{Acknowledgments}
YBO gratefully acknowledges financial support from Consejo Nacional de Ciencia y Tecnolog\'ia de M\'exico (CONACYT, M\'exico) Proyecto Fronteras 952 and the UNAM-PAPIIT research Projects No. IN-103017 and No. IA-101618. EDB acknowledges support from UPIIH of the Instituto Polit\'ecnico Nacional (IPN, M\'exico). The authors also thank Thomas H. Seligman, Thomas Stegmann, Francois Leyvraz, Carlos A. Gonz\'alez-Guti\'errez, Francisco J. Turrubiates, and Maurice Oliva-Leyva for helpful discussions, comments, and criticisms.

\appendix
\numberwithin{equation}{section}
\section{Wigner function: the Moyal star product}\label{A}

In this appendix, we present an alternative procedure to obtain the WF for the DW problem.

As was shown in the above sections, the WF $W(\vec{r},\vec{p})$ is a real function defined in phase space and deformation quantization formulation (DQF) provides an alternative method to obtain it. The implementation of the DQF on the study of one-dimensional quantum systems has allowed us, in principle, to recover some previous results obtained from canonical quantization but described in phase space.

In the QDF, the eigenvalue equation $H\psi=E\psi$ is replaced by a $\star$-genvalue equation, namely:
\begin{equation}
H\star W=W\star H=EW,
\end{equation}
for describing the eigenvalue energy $E$, where the star product ($\star$) represents a deformation performed on the algebraic structure in phase space of classical mechanics, which carries a noncommutative algebra in that of quantum mechanics.

One of the star products defined in the DQF framework is the so-called Moyal star product~($\star_{M}$) \cite{zachos02,hh02}:
\begin{equation}
f\star_{M}g\equiv f\exp\left(\frac{i\hbar}{2}\left(\overleftarrow{\partial_{\vec{r}}}\overrightarrow{\partial_{\vec{p}}}+\overleftarrow{\partial_{\vec{p}}}\overrightarrow{\partial_{\vec{r}}}\right)\right)g,
\end{equation}
where $\overleftarrow{\partial}$ ($\overrightarrow{\partial}$) indicates that the derivative acts on the function to the left (right). Alternatively, the Moyal star product can be rewritten as 
\begin{subequations}\label{bopp}
	\begin{align}
	f(\vec{r},\vec{p})\star_{M}g(\vec{r},\vec{p})&=f\left(\vec{r}+\frac{i\hbar}{2}\overrightarrow{\partial_{\vec{p}}},\vec{p}-\frac{i\hbar}{2}\overrightarrow{\partial_{\vec{r}}}\right)g(\vec{r},\vec{p}), \label{5a} \\
	f(\vec{r},\vec{p})\star_{M}g(\vec{r},\vec{p})&=f(\vec{r},\vec{p})g\left(\vec{r}-\frac{i\hbar}{2}\overleftarrow{\partial_{\vec{p}}},\vec{p}+\frac{i\hbar}{2}\overleftarrow{\partial_{\vec{r}}}\right). \label{5b}
	\end{align}
\end{subequations}
Thus, the WF satisfies the $\star_{M}$-genvalue equations
\begin{subequations}
	\begin{align}
	H(\vec{r},\vec{p})\star_{M}W(\vec{r},\vec{p})&=H\left(\vec{r}+\frac{i\hbar}{2}\overrightarrow{\partial_{\vec{p}}},\vec{p}-\frac{i\hbar}{2}\overrightarrow{\partial_{\vec{r}}}\right)W(\vec{r},\vec{p})=EW(\vec{r},\vec{p}), \\
	H(\vec{r},\vec{p})\star_{M}W(\vec{r},\vec{p})&=W(\vec{r},\vec{p})H\left(\vec{r}-\frac{i\hbar}{2}\overleftarrow{\partial_{\vec{p}}},\vec{p}+\frac{i\hbar}{2}\overleftarrow{\partial_{\vec{r}}}\right)=EW(\vec{r},\vec{p}),
	\end{align}
\end{subequations}
where $E$ is the energy eigenvalue of $H\psi=E\psi$.

According to~\cite{Yuan}, we can solve the Dirac-Weyl equation~(\ref{WDE}) by applying the Moyal star product:
\begin{equation}\label{24}
H_{\rm D}(\vec{r},\vec{p})\star W(\vec{r},\vec{p})=
EW(\vec{r},\vec{p}),
\end{equation}
where
\begin{eqnarray}\label{16}
W(\vec{r},\vec{p})=\left(\begin{array}{c c}
W^{(a)}(\vec{r},\vec{p}) & 0 \\
0 & W^{(b)}(\vec{r},\vec{p})
\end{array}\right), \quad W^{(a,b)}(\vec{r},\vec{p})=\left(\begin{array}{c c}
W_{11}^{(a,b)}(\vec{r},\vec{p}) & W_{12}^{(a,b)}(\vec{r},\vec{p}) \\
W_{21}^{(a,b)}(\vec{r},\vec{p}) & W_{22}^{(a,b)}(\vec{r},\vec{p})
\end{array}\right).
\end{eqnarray}
This means that the WF is a $4\times4$ matrix but that in an adequate representation it can be reduced to a diagonal partitioned matrix, as is shown in Eq.~(\ref{16})~\cite{Yuan}. Here, the $2\times2$ WM $W^{(a)}(\vec{r},\vec{p})$ corresponds to Dirac fermions at the $K_D$ valley, while $W^{(b)}(\vec{r},\vec{p})$ corresponds to those at the $K'_D$ valley, so that we will just focus on the former. 

Equation~(\ref{24}) gives rise to two decoupled equations:
\begin{subequations}
	\begin{align}
	v_{\rm F}\,\left(a\sigma_{x}\left[p_x+eA_x\right]+b\sigma_{y}\left[p_y+eA_y\right]\right)\star  W^{(b)}&=EW^{(a)}, \\ v_{\rm F}\,\left(a\sigma_{x}\left[p_x+eA_x\right]+b\sigma_{y}\left[p_y+eA_y\right]\right)\star  W^{(a)}&=EW^{(b)}.
	\end{align} 
\end{subequations}
After decoupling the above expressions, we obtain the following $\star$-genvalue equation for $W^{(a)}(\vec{r},\vec{p})$:
\begin{align}\label{50}
\nonumber	H_{\rm D}^{(a)}\star W^{(a)}&=\left[	v_{\rm F}\,\left(a\sigma_{x}\left[p_x+eA_x\right]+b\sigma_{y}\left[p_y+eA_y\right]\right)\right]\star \\
&\quad\star\left[	v_{\rm F}\,\left(a\sigma_{x}\left[p_x+eA_x\right]+b\sigma_{y}\left[p_y+eA_y\right]\right)\right]\star W^{(a)}=E^2W^{(a)},
\end{align}
where $[W^{(a)}(\vec{r},\vec{p})]_{ij}\equiv W_{ij}(x,p_{x})W_{ij}(y,p_{y})$, $i,j=1,2$.

Considering the Landau gauge $\vec{A}=B_0x\hat{e}_y$ and Bopp's shift (Eq.~(\ref{5a})), the Hamiltonian $H_{\rm D}^{(a)}$ yields:
\begin{align}\label{hamiltonian}
\nonumber H_{\rm D}^{(a)}&=v_{\rm F}^2\hbar^2\left[a\sigma_{x}\left(\frac{p_x}{\hbar}-\frac{i}{2}\partial_x\right)+b\sigma_{y}\left[\left(\frac{p_y}{\hbar}-\frac{i}{2}\partial_y\right)+\frac{eB_0}{\hbar}\left(x+\frac{i\hbar}{2}\partial_{p_x}\right)\right]\right]^2 \\
\nonumber&=abv_{\rm F}^2\hbar^2\left\{\left[\zeta\left(\frac{p_x^2}{\hbar^2}-\frac{1}{4}\partial_x^2\right)+\zeta^{-1}\left(\frac{p_y}{\hbar}+\frac{\omega_{\rm B}}{2}x\right)^2-\frac{1}{4}\zeta^{-1}\left(\frac{\omega_{\rm B}\hbar}{2}\partial_{p_x}-\partial_y\right)^2\right]\mathbb{I}+\frac{\omega_{\rm B}}{2}\sigma_{z}\right.\\
&\quad\left.+i\left[-\zeta\frac{p_x}{\hbar}\partial_x+\zeta^{-1}\left(\frac{p_y}{\hbar}+\frac{\omega_{\rm B}}{2}x\right)\left(\frac{\omega_{\rm B}\hbar}{2}\partial_{p_x}-\partial_y\right)\right]\mathbb{I}\right\},
\end{align}
where $\mathbb{I}$ is the $2\times2$ identity matrix, $\omega_{\rm B}=\zeta\omega_{\zeta}$ and $\sigma_{z}$ is the third Pauli matrix. Here, we could assume that the corresponding pseudo-spinor is an eigenstate of the operator $\sigma_{z}$ in order to reduce the WF to a diagonal matrix. However, we will work with the entire representation.

By defining the quantities
%
\begin{equation}\label{A9}
\mathcal{E}_1=\frac{E^2}{abv_{\rm F}^2\hbar^2}-\frac{\omega_{\rm B}}{2}, \quad \mathcal{E}_2=\frac{E^2}{abv_{\rm F}^2\hbar^2}+\frac{\omega_{\rm B}}{2},
\end{equation}
and using the variables $\xi$ and $s$ defined in (\ref{19}), we can obtain the following pair of Hamiltonians:
\begin{subequations}\label{hamiltonians}
	\begin{align}
	H_1&=\frac{\omega_{\rm B}}{2}\left[s^2+\left(\sqrt{\frac{2}{\zeta\omega_{\rm B}}}\left(\frac{p_y}{\hbar}-k\right)+\xi\right)^2-\frac{1}{4}\partial_{\xi}^2-\frac{1}{4}\left(\partial_{s}-\sqrt{\frac{2}{\zeta\omega_{\rm B}}}\partial_y\right)^2\right], \\
	H_2&=\frac{\omega_{\rm B}}{2}\left[-s\partial_{\xi}+\left(\sqrt{\frac{2}{\zeta\omega_{\rm B}}}\left(\frac{p_y}{\hbar}-k\right)+\xi\right)\left(\partial_{s}-\sqrt{\frac{2}{\zeta\omega_{\rm B}}}\partial_y\right)\right].
	\end{align}
\end{subequations}

Following the ansatz shown in Ref.~\cite{Kryukov} for the WF $W_{ij}(y,p_{y})$, the solutions for the problem in Eq.~(\ref{50}) are given as
\begin{subequations}
	\begin{align}
	W(p_{y})&\equiv W_{ij}(y,p_{y})=\delta(p_{y}-k\hbar), \quad i,j=1,2, \\
	W_{1j}(x,p_{x})&=\frac{(-1)^{n-1}}{\pi}\exp\left(-\frac{1}{2}\vert\chi\vert^2\right)L_{n-1}\left(\vert\chi\vert^2\right), \quad j=1,2, \\
	W_{2j}(x,p_{x})&=\frac{(-1)^{n}}{\pi}\exp\left(-\frac{1}{2}\vert\chi\vert^2\right)L_{n}\left(\vert\chi\vert^2\right), \quad j=1,2,
	\end{align}
\end{subequations}
where again $\chi=\sqrt{2}(\xi+is)$, and the corresponding energy spectrum turns out to be
\begin{equation}\label{energy}
E_{2,0}=0, \quad E_{2,n}=E_{1,n-1}=\textrm{sgn}(n) v'_{\rm F}\hbar\sqrt{\omega_{\rm B}\vert n\vert},
\end{equation}
$n=0,\pm1,\pm2,\dots$, sgn$(0)=1$ and $v'_{\rm F}=\sqrt{ab}\,v_{\rm F}$.~Meanwhile, the $2\times2$ Wigner matrices in~(\ref{16}) can be rewritten as
\begin{equation}\label{mwigner}
W^{(a,b)}(\vec{r},\vec{p})=\frac{1}{2^{(1-\delta_{0n})}}\left(\begin{array}{c c}
W_{11}^{(a,b)}(\vec{r},\vec{p}) & W_{12}^{(a,b)}(\vec{r},\vec{p}) \\
W_{21}^{(a,b)}(\vec{r},\vec{p}) & W_{22}^{(a,b)}(\vec{r},\vec{p})
\end{array}\right),
\end{equation}
where the components are the following WFs:
\begin{subequations}\label{wigner}
	\begin{align}
	W_{11}^{(a,b)}(\vec{r},\vec{p})&=W_{12}^{(a,b)}(\vec{r},\vec{p})=(1-\delta_{0n})\frac{(-1)^{n-1}}{\pi}\delta\left(p_y-k\hbar\right)\exp\left(-\frac{1}{2}\vert\chi\vert^2\right)L_{n-1}\left(\vert\chi\vert^2\right), \label{A17} \\
	W_{21}^{(a,b)}(\vec{r},\vec{p})&=W_{22}^{(a,b)}(\vec{r},\vec{p})=\frac{(-1)^{n}}{\pi}\delta\left(p_y-k\hbar\right)\exp\left(-\frac{1}{2}\vert\chi\vert^2\right)L_{n}\left(\vert\chi\vert^2\right), \label{A18}
	\end{align}
\end{subequations}
such that
\begin{equation}
\int_{-\infty}^{\infty}\int_{-\infty}^{\infty}\text{Tr}\left[W^{(a,b)}(\vec{r},\vec{p})\right]d\vec{r}d\vec{p}=1.
\end{equation}
Notice that the factor $(1-\delta_{0n})$ in Eqs.~(\ref{mwigner}) and (\ref{A17}) guarantees that such expressions satisfy the $\star$-genvalue equation~(\ref{50}) for the Landau level $n=0$.

\subsection{Comparison with the integral representation}
Coming back to the WF obtained by the integral representation (Eq.~(\ref{6})), we can identify each component in Eq.~(\ref{23}) with the given ones in (\ref{wigner}), except the off-diagonal terms:
\begin{equation}
W_{12}^{(a,b)}(\vec{r},\vec{p})\neq W_{n-1,n}(\vec{r},\vec{p}), \quad W_{21}^{\ast(a,b)}(\vec{r},\vec{p})\neq W_{n,n-1}(\vec{r},\vec{p}).
\end{equation}
This is because in both representations the trace of the WF is an invariant, while the off-diagonal terms depend on the representation considered.

However, we can establish a connection between the two representations as follows. By acting the Hamiltonians in Eq.~(\ref{hamiltonians}) on the above expressions and recalling that $f(x)\delta(x-a)=f(a)\delta(x-a)$, we get:
\begin{subequations}\label{61}
	\begin{align}
	\left(H_1+\frac{\omega_{\rm B}}{2}\right)W_{n-1,n}(\vec{r},\vec{p})&=\left(n+\frac12\right)\omega_{\rm B} W_{n-1,n}(\vec{r},\vec{p}), \quad H_2W_{n-1,n}(\vec{r},\vec{p})=i\,\frac{\omega_{\rm B}}{2}W_{n-1,n}(\vec{r},\vec{p}), \label{A17a} \\
	\left(H_1-\frac{\omega_{\rm B}}{2}\right)W_{n,n-1}(\vec{r},\vec{p})&=\left(n-\frac12\right)\omega_{\rm B} W_{n,n-1}(\vec{r},\vec{p}), \quad H_2W_{n,n-1}(\vec{r},\vec{p})=-i\,\frac{\omega_{\rm B}}{2}W_{n,n-1}(\vec{r},\vec{p}). \label{A17b}
	\end{align}
\end{subequations}
By taking the complex conjugate of Eqs.~\eqref{A17b} and adding them to~\eqref{A17a}, we have that
\begin{equation}
H_1\Re\left[W_{n-1,n}(\vec{r},\vec{p})\right]=n\omega_{\rm B}\Re\left[W_{n-1,n}(\vec{r},\vec{p})\right], \quad H_2\Re\left[W_{n-1,n}(\vec{r},\vec{p})\right]=0.
\end{equation}
Therefore, although separately the off-diagonal terms of the WF $W_{n}(\vec{r},\vec{p})$ in Eq.~(\ref{6}) are not solutions of the $\star$-genvalue equation~(\ref{50}), their real part is. Thus, the integral representation of the WF also gives rise to a matrix function but in another representation.

\bibliographystyle{ieeetr}
\bibliography{biblio2_1}

\end{document}